\documentclass[amsmath,amssymb,onecolumn,superscriptaddress]{revtex4}

\usepackage{graphicx}%
\usepackage{dcolumn}%
\def\be{\begin{equation}}
\def\ee{\end{equation}}
\def\beq{\begin{eqnarray}}
\def\eeq{\end{eqnarray}}

\usepackage{bm}
\usepackage{graphicx}
\usepackage{dcolumn}
\usepackage{amsmath}
\usepackage[latin1]{inputenc}
\usepackage{graphicx, psfrag}
\usepackage{amssymb}
\usepackage[colorlinks=true, citecolor=blue, urlcolor = blue, linkcolor= red, bookmarks=true]{hyperref}
\usepackage{float}
\usepackage{mathtools}
\usepackage{amsmath}
\usepackage{amsfonts}
\usepackage{dcolumn}
\usepackage{hyperref}
\usepackage{subfigure}
\usepackage{pgfplots}
\usepackage{epstopdf}
\usepackage{booktabs}
\usepackage{orcidlink}
\usepackage{xcolor}

\begin{document}
\title{Celestial attributes of Hybrid star in $5\mathcal{D}$ Einstein-Gauss-Bonnet gravity}

\author{Akashdip Karmakar \orcidlink{0009-0007-3848-1443}}
\email[Email:]{akashdip999@gmail.com}
 \affiliation{Department of Mathematics,  Indian Institute of Engineering Science and Technology, Shibpur, Howrah, West Bengal 711 103, India}
 
\author{Pramit Rej \footnote{Corresponding author} \orcidlink{0000-0001-5359-0655}}
\email[Email:]{pramitrej@gmail.com, pramitr@sccollegednk.ac.in, pramit.rej@associates.iucaa.in}
 \affiliation{Department of Mathematics, Sarat Centenary College, Dhaniakhali, Hooghly, West Bengal 712 302, India}

\begin{abstract}
Hybrid star is the term given to a neutron star with a quark core. Due to a lot of uncertainties in the calculations and compositions of such a high-density system, it is of great interest and a preferred scenario for particle physicists and astrophysicists. To explore some novel aspects within the framework of the $5\mathcal{D}$ Einstein-Gauss-Bonnet(EGB) gravity, our current study presents a hybrid star model that includes strange quark matter in addition to regular baryonic matter. In hybrid stellar objects, a hadronic outer component surrounds a quark inner component, which prompts the consideration of the most basic MIT bag model equation of state to correlate the density and pressure of strange quark matter within the stellar interior, while radial pressure and matter density due to baryonic matter are connected by a linear equation of state. The model is constructed within the specifications of the Krori and Barua (KB) {\em ansatz} (Krori and Barua, J. Phys. A: Math. Gen. $\bold{8},508, 1975$). Here we present the solution for a particular compact object 4U 1538 - 52 with mass $\mathcal{M} = 0.87 \pm 0.07~\mathcal{M}_{\odot}$ and radius $\mathfrak{R} = 7.866_{-0.21}^{+0.21}$ km. We examine the fundamental physical characteristics of the star, which highlights how the values of matter variables are affected by the Gauss-Bonnet coupling parameter $\alpha$. Finally, as it meets all the physical requirements for a realistic model, we have come to realize that our present model is realistic.
\end{abstract}

\maketitle

\section{Introduction}

Among physicists and theoretical academics, there has been an incredible level of inquiry over the topic of cosmic accelerating expansion and the nature of dark energy(DE). Several attempts have been made for this goal in various mathematical contexts. These efforts can be divided into two categories: modifying the Einstein-Hilbert action to produce alternative theories of gravity, and proposing new DE parameters for the entire cosmic evolution \cite{abbas2021hybrid}. The aim of the investigation topic would be to evaluate the physical characteristics of relativistic compact objects like white dwarfs, neutron stars, pulsars, quark-made stars, and black holes in contrast to conventional theories of gravity. The specific characteristics of the compact objects, which are generated at the highest stages of star evolution, are yet unclear, although numerous indications point to their extreme mass and small radius. It is also to be mentioned that compact objects are divided into various groups based on the mass-to-radius ratio, including white dwarfs, neutron stars, ultra-compact objects, black holes, etc \cite{jotania2006paraboloidal}. The study of a huge gravitational field in dense compact objects is a perfect example of how to describe the differences between General Relativity (GR) and its modification. However, there are a number of theoretical and observational issues with GR that remain unresolved, leading to the development of several alternative models. Numerous modified theories of gravity have been recorded in literature, covering a variety of topics\cite{deb2019study, ilyas2018charged, shahzad2019strange, bhatti2021dynamical, pretel2022electrically, panotopoulos2022charged, goswami2014collapsing, chilambwe2015new, hansraj2019constant,hansraj2022strange, govender2019role, bhar2022tolman, rej2023isotropic, maurya2022role, errehymy2022anisotropic, ditta2022physical, rahaman2020anisotropic, maurya2020anisotropic, singh2020exploring, singh2020physical, rej2022model, Salti1, Salti2}. The higher-curvature gravity theory proposed by Lovelock (sometimes referred to as Lanczos-Lovelock) is somewhat unique in this context \cite{lovelock1971einstein, lovelock1972four}. Lovelock theory, specifically, extends GR in higher-dimensional spacetimes while retaining the field equations' order down at second order in derivatives without torsion. Furthermore, the standard formulations of string theory call for a total of 10 dimensions, or 11 when we include an extended formulation known as M-Theory i.e. the existence of extra spacetime dimensions beyond the four \cite{kaluza2018unification, klein1926quantentheorie}. The EGB gravity, which has a lagrangian that is the sum of a cosmological constant and a curvature scalar, and a third term that contains a quadratic Gauss-Bonnet (GB) term, is the simplest non-trivial Lovelock gravity among all the classes. The required second-order equations of motion are produced by the GB Lagrangian in an N-dimensional spacetime (where $N \ge 5$). Higher dimensions are a logical extension of the gravitational field's behavior and dynamics. The so-called EGB gravity, which results from the addition of an extra term to the basic Einstein-Hilbert action and is quadratic in the Riemann tensor, is a useful generalization of classical general relativity.
Only a system of second-order equations that are consistent with classical general relativity results from varying this additional factor with regard to the metric tensor \cite{bhar2017comparative}. Another intriguing finding was that stability may be attained with more mass in $5\mathcal{D}$ EGB than in conventional $4\mathcal{D}$ Einstein gravity \cite{goswami2015buchdahl}. Similar to the $4\mathcal{D}$ scenario, the $5\mathcal{D}$ EGB models demonstrate a linear barotropic equation of state \cite{hansraj2015exact, maharaj2015exact}.
\\
Because it is used to describe how compact things behave, finding a specific solution to the Einstein field equations is additionally significant. The most advantageous state of baryon matter is called strange quark matter, which is made up of u, d, and s quarks. Since a baryon is made up of a number of elementary particles, it is attainable that unbound basic particles might be present inside superdense stars, which leads to the existence of the quark star model \cite{itoh1970hydrostatic}.
When matter density of a star reaches extremely high levels towards its center and the star core contains quark matter, the de-confinement transition takes place. The term ``Hybrid stars" refers to these compact objects. In our universe, there are essentially two different sorts of compact stars.
Quark or hybrid neutron-quark stars are made up entirely or mostly of quark matter. The first is referred to as a strange quark star(SQS), while the second is referred to as a hybrid star. Numerous studies have been conducted to show the likely existence of quark matter(QM) in massive Neutron stars(NS)\cite{buballa2014emmi, weber2014properties, orsaria2014quark, drago2014can, alvarez2016new, alford2016characteristics}.

The quark star and the hybrid star take up a significant amount of space among several fictitious compact objects that have been the subject of literature discussion since they are gravitationally wave-producing things. Moreover, gravitational wave detection techniques open up the prospect of determining these unique stars. Due to the fact that strange stars can release gravitational wave echoes, it is acceptable to acquire more about the equation of state (EoS) of quark matter by studying the gravitational waves they emit \cite{mannarelli2018gravitational, sotani2004restricting}. Hadronquark transition can be described in a variety of ways, some of which have already been explored in previous studies \cite{benic2015effective, marczenko2018chiral, masuda2013hadron, kovacs2020hybrid}.
In order to characterize strange matter inside strange stars, the MIT bag model was developed, where color confinement has been proposed to be a bag, and a bag constant ($B_g$) is taken into consideration in order to determine the size of the bag \cite{farhi1984strange}. Based on the strange quark mass, the $B_g$ can be determined and it can be considered to be a gap between the energy densities of true and false vacuum. Also, the energy density can be obtained by adding $B_g$ to the kinetic energy of the quarks \cite{bordbar2012calculation}. The pressure-density relation
for QM is given by using the MIT bag EoS as: $p_q = \frac{1}{3}(\rho_q - 4B_g)$, where $\rho_q$ and
$p_q$ are, respectively, the density and pressure for QM. It is also preferable to include an equation of state linking the density($\rho$) to the pressure profile($p_r$) in stellar modeling. In this paper, we have considered: $p_r = \beta \rho + \gamma$.\\
Rahaman and his co-workers have identified a background space-time metric for the galactic halo region by employing the flat rotation curve as input and treating the matter content as quark matter \cite{rahaman2012quark}.
Witten identified two mechanisms for the creation of strange matter: the quark-hadron phase transition in the early cosmos and the transformation of neutron stars into strange ones at extremely high densities \cite{witten1984cosmic}. Bodmer also theorized that strange quark matter is likely to exist in the inner core of the neutron star and that a phase transition between hadronic and strange quark matter might occur in the universe when a large star breaks down as a supernova with a density higher than nuclear density \cite{bodmer1971collapsed}.\\
In the current work, we present a hybrid star model with regular baryonic matter together with strange quark matter(SQM) in EGB employing two different forms of EoSs, which are mentioned previously. The model is developed by utilizing KB {\em ansatz} \cite{krori1975singularity}. The rest of the parts of the paper are arranged as follows. We covered the inner spacetime in Section \ref{Sec2} along with the fundamental field equations. In the next section, we have discussed stellar configuration, i.e. exterior spacetime, equation of state, and solutions of the system. In this Section, the values of the unknown parameters have been obtained by smoothly matching our inner spacetime to the outer Schwarzschild line element. We have covered the regularity of the metric coefficients, density, pressure, redshift, and the EoS parameter for hybrid star models in Section \ref{Sec4}, which is devoted to the physical evaluation of the results. In section \ref{Sec5}, we have checked the stability of our proposed model. In Section \ref{Sec6}, we have finally compiled every aspect of our work for the current star model.

\section{Preliminaries: theoretical setup}\label{Sec2}
We require an adjusted action that is different from the Einstein situation in order to derive the field equations in EGB gravity.
Now, The following total action is used to generate the $\mathcal{ D}(\geq 5)$-dimensional formulation of the EGB gravity\cite{maeda2008generalized} i.e.
\begin{eqnarray}\label{action}
\mathcal{S_T} = \frac{1}{2 {\kappa} }\int d^{\mathcal{D}}x\sqrt{-g}\left[ R-2\Lambda +\alpha \mathcal{L}_{\text{GB}} \right]+ \mathcal{S}_{\text{matter}},    
\end{eqnarray}
where ``$g$" denotes the determinant of the metric tensor ``$g_{ij}$", ``$R$" shows the Ricci curvature scalar, and ``$\Lambda$" represents the commonly recognized cosmological constant, the Lagrangian density $\mathcal{L}_{\text{GB}}$ refers to the GB term, $\mathcal{S}_{\text{matter}}$ represents the action relating to the matter field, and $\alpha$ is a free coupling parameter with the dimension [length]$^2$. Now, we are to focus on five dimensions in this study.\\
It is important to keep in mind that the auxiliary coupling constant, which grows with the length-square dimension and only accepts nonnegative values (\cite{Maeda2006final}), is associated with the string tension in string theory and reflects the UV corrections to Einstein's theory of the GR \cite{Sepul}.\\
Additionally, the expression below defines the GB word $\mathcal{L_\text{GB}}$: 
\begin{equation}
\mathcal{L}_{\text{GB}}=R^{ijkl} R_{ijkl}- 4 R^{ij}R_{ij}+ R^2\label{GB}
\end{equation}
and the stress-energy tensor $T_{ij}$ for the matter field is set off by the matter portion of the total action, or $\mathcal{S}_\text{matter}$. Therefore, modifying the $\mathcal{D}=5$ scenario of the aforementioned action in regard to the $g_{ij}$ metric tensor results in the following equations of motion.

\begin{equation}\label{eq3}
G_{ij}+\alpha H_{ij} = \kappa  T_{ij}, 
\end{equation}
where
\begin{eqnarray}
G_{ij} = R_{ij}-\frac{1}{2}R~ g_{ij},
\end{eqnarray}
\begin{eqnarray}
H_{ij} =  2\Big( R R_{ij}-2R_{ik} {R}^k_j -2 R_{ijkl}{R}^{kl} - R_{ikl\delta}{R}^{kl\delta}_j\Big) - \frac{1}{2}~g_{ij}~\mathcal{L}_{\text{GB}},
\end{eqnarray}
and
\begin{equation}\label{eq3b}
{T}_{ij}=-\frac{2}{\sqrt{-g}}\frac{\delta\left(\sqrt{-g}\mathcal{S}_m\right)}{\delta g^{ij}}. 
\end{equation}
The energy-momentum tensor ${T}_{ij}$ corresponding to the matter field is obtained from $\mathcal{S}_{\text{matter}}$.
In the current article, we have considered a model of a hybrid star that consists of both ordinary baryonic matter with density $\rho$ and strange quark matter with density $\rho_q$ and for the purpose of simplicity, we have not taken into account the interaction between these two materials. 
So, the energy-momentum tensor for the stellar fluid in our model is taken to be
\begin{equation}\label{eq3i}
{T}_{ij}=diag(-\rho^{eff},p_r^{eff},p_t^{eff},p_t^{eff},p_t^{eff}),
\end{equation}

where $\rho^{eff} = (\rho + \rho_q)$, $p_r^{eff} = (p_r + p_q)$ and $p_t^{eff} = (p_t + p_q)$.\\
Here $\rho$, $p_r$, and $p_t$ respectively denote the matter-energy density, radial pressure, and tangential pressure of the baryonic matter, whereas $\rho_q$ and $p_q$ denote the respective matter density and pressure due to quark matter.\\

We can now think about using a static, spherically symmetric $5\mathcal{D}$ geometry in the governing equations of the EGB framework. Therefore, we suppose that the following line element in coordinates $(x^i = t, r, \theta, \phi, \psi)$. serves as a representation of the interior of our celestial object i.e.

\begin{eqnarray}
\label{5} ds^{2}_{\bold{int}}= -e^{2\nu(r)} dt^{2} + e^{2\lambda(r)} dr^{2} +
 r^{2}(d\theta^{2} + \sin^{2}{\theta} d\phi^2 +\sin^{2}{\theta} \sin^{2}{\phi}
 d\psi^2),
\end{eqnarray}
where $e^{\nu}$ and $e^{\lambda}$ encipher the characteristics of the gravitational field that depend on $r$ only. \\
As a result, the remaining parts of the metric tensor $g_{ij}$ and its inverse $g^{ij}$ are as follows:
\begin{eqnarray}\label{5b} 
g_{ij}=-e^{2\nu}\delta_i^0\delta_j^0+e^{2\lambda}\delta_i^1\delta_j^1+r^2\delta_i^2\delta_j^2+r^2\sin^{2}{\theta}(\delta_i^3\delta_j^3+\sin^{2}{\phi}\delta_i^5\delta_j^5),
\end{eqnarray}
\begin{eqnarray}\label{5c} 
g^{ij}=e^{-2\nu}\delta^i_0\delta^j_0+e^{-2\lambda}\delta^i_1\delta^j_1+\frac{1}{r^2}\delta^i_2\delta^j_2+\frac{1}{r^2 \sin^{2}{\theta}}\left(\delta^i_3\delta^j_3+\frac{1}{\sin^{2}{\phi}}\delta^i_5\delta^j_5\right).
\end{eqnarray}
In light of the metric above, the following set of independent equations can be obtained by assuming that the five-velocity components are specified by $u^i=e^{-\nu}\delta_0^i$ in the EGB system.

\begin{eqnarray}
\label{7a}\kappa\rho^{eff}= \kappa(\rho + \rho_q) &=& -\frac{3}{e^{4\lambda }r^3} \Bigg[re^{2\lambda}- r^2 e^{2\lambda}\lambda'  -4\alpha e^{2\lambda}\lambda' -re^{4\lambda} + 4\alpha \lambda'\Bigg],\label{fe1}\\
\label{7b} \kappa p_r^{eff} = \kappa(p_r + p_q) & = &\frac{3}{e^{4\lambda }r^3}
\Big[(r^2 \nu' +r +4\alpha \nu')e^{2\lambda} -re^{4\lambda} -4\alpha \nu'\Big] , \label{fe2}\\
\label{7c} \kappa p_t^{eff} = \kappa(p_t + p_q) &=& 
 \frac{1}{e^{2\lambda }r^2} \Big[1 +2r \nu'
-2r\lambda' - r^2\nu'\lambda' + r^2(\nu')^2 \Big] +\frac{1}{e^{2\lambda}r^2} \Big[ -4\alpha\nu'\lambda' + 4\alpha (\nu')^2  +4\alpha \nu'' + r^2 \nu''\Big] \nonumber \\ && + \frac{1}{e^{4\lambda }r^2} \Big[12 \alpha \nu' \lambda' -4 \alpha(\nu')^2 - 4\alpha \nu'' - e^{4\lambda}\Big]\label{fe3}
\end{eqnarray}
where the prime $(')$ symbol denotes the differentiation with respect to the radial coordinate `$r$'.

\section{Stellar configuration}\label{Sec3}

Fundamental junction conditions promise harmonious matching at the boundary for the exterior and interior solutions of a static stellar object and in addition, even though the surface of a celestial structure should not have any radial pressure, the boundary does not always lose tangential pressure. These lead to investigate in detail the junction conditions of a Hybrid star.
To characterize the inner region of a compact star structure, many techniques have been provided in the literature. To construct a stellar model, we have considered KB {\em ansatz} \cite{Krori}given by \\ $2\nu=Br^2+C$ and $2\lambda=Ar^2$,\\
where $A$, $B$, and $C$ represent arbitrary constant parameters. It is important to smoothly match the interior space-time solution with a suitable static and spherically symmetric exterior vacuum Schwarzschild formulation in order to derive the constant parameters, i.e., $A$, $B$, and $C$ for our suggested model. Moreover, the Schwarzschild vacuum solution is essential to astrophysics because of its asymptotically flat nature.
The EGB version of the Schwarzschild (EGB-Sch) metric (or the Boulware-Deser geometry) is the most acceptable exterior vacuum solution since we are working with the interior of an anisotropic quark star in the $5\mathcal{D}$ EGB formalism and it is given by  
\cite{boulware1985string, rej2023charged}

\begin{eqnarray}\label{ext}
ds^{2}_{\bold{ext}} &=& -\mathcal{F}(r)dt^2 + \frac{dr^2}{\mathcal{F}(r)} + r^2(d\theta^2+\sin^2\theta d\phi^2 + \sin^2\theta\sin^2\phi^2 d\psi^2), \label{metric}
\end{eqnarray}
where 
\begin{eqnarray}
\mathcal{F}(r) = K + \frac{r^2}{4\alpha}\left(1 - \sqrt{1 + \frac{8\alpha \mathcal{M}}{r^4}}\right), 
\end{eqnarray}
where $K$ denotes an arbitrary constant, and $\mathcal{M}$ indicates the gravitational mass of the compact star. We assert that in the case of a hybrid quark star, the intrinsic solution of the star relates perfectly with the EGB-Sch solution, allowing us to compare the interior metric to the EGB-Sch exterior vacuum spacetime.
So, the Krori-Barua configuration method is the next thing we look at in our investigation such as:

\begin{equation}\label{conf1}
e^{2\nu(r)}=\left\{\begin{array}{ll}
                    e^{Br^2 + C} & \qquad r<{\mathfrak{R}}, \\
                    K + \frac{r^2}{4\alpha}\left(1 - \sqrt{1 + \frac{8\alpha \mathcal{M}}{r^4}}\right) & \qquad r>{\mathfrak{R}},
                  \end{array}
\right.
\end{equation}
\begin{equation}\label{conf2}
e^{2\lambda(r)}=\left\{\begin{array}{ll}
                    e^{Ar^2} & \qquad r<{\mathfrak{R}}, \\
                    \left[K + \frac{r^2}{4\alpha}\left(1 - \sqrt{1 + \frac{8\alpha \mathcal{M}}{r^4}}\right)\right]^{-1} & \qquad r>{\mathfrak{R}}.
                  \end{array}
\right.
\end{equation}
Note that, in the chosen configuration strategy, (i) {$\mathfrak{R}$} is the radius of the star with unit in $\rm{km}$, (ii) $B$ and $A$ are constants with units respectively in $\rm{km}^{-2}$and $\rm{km}^{-2}$, (iii) $C$ is a dimensionless parameter. The numerical values of these constants will be determined by a smooth matching of the chosen interior and outside solutions.
With the help of the expressions given in (\ref{conf1}) and (\ref{conf2}), the field equations (\ref{fe1})-(\ref{fe3}) take the following forms for the interior part of the star:
\begin{eqnarray}
\kappa(\rho + \rho_q)  &=& \frac{3e^{-2Ar^2}}{r^2}\Bigg[-4A\alpha + e^{2Ar^2} + e^{Ar^2}\{-1 + A(4\alpha + r^2)\}\Bigg], \label{f1} \\
\kappa(p_r + p_q) &=& \frac{3e^{-2Ar^2}}{r^2}\Bigg[-4\alpha B - e^{2Ar^2} + e^{Ar^2}(1 + 4\alpha B + Br^2)\Bigg], \label{f2} \\
\kappa(p_t + p_q) &=& \frac{e^{-2Ar^2}}{r^2}\Bigg[-e^{2Ar^2} - 4\alpha B\{1 + (-3A + B)r^2\} + e^{Ar^2}\Big[1 + 4\alpha B + \{-2A(1 + 2\alpha B) + \nonumber\\ &&  B(3 + 4\alpha B)\}r^2 + B(-A + B)r^4\Big]\Bigg].\label{f3}
 \end{eqnarray}

\subsection{Assumption of Equation of State due to the quark matter}\label{eosq}

Although the system of equations is mathematically well stated, it is too difficult to acquire the explicit solutions to the aforementioned field equations (\ref{f1}-\ref{f3}).
Therefore, we have implemented two assumptions to eliminate this complexity.\\
(i) We have selected the linear equation of state given by 
\begin{equation}\label{eos1}
p_r = \beta \rho + \gamma,    
\end{equation}
i.e. a convincing relationship arises between the density of normal baryonic matter $(\rho)$ and the radial pressure $p_r$. where $0 < \beta < 1$ with $\beta \neq \frac{1}{3}$. and $\gamma > 0$.\\

(ii) Assuming further that the pressure-matter density relation for quark matter is given by the MIT bag model equation of state as follows:
\begin{equation}\label{eos2}
    p_q  = \frac{1}{3}(\rho_q - 4B_g)
\end{equation}
where $B_g$ is the bag constant of unit $\rm{MeV}/\rm{fm^3}$.

\subsection{Proposed stellar model with baryonic and strange quark matter}

 Now, solving the equations (\ref{f1}), (\ref{f2}) and (\ref{f3}) with the help of (\ref{eos1}) and  (\ref{eos2}), we get the expressions for the matter-energy density, radial pressure, transverse pressure for our model as:
\begin{eqnarray}
\rho &=& \frac{e^{-2Ar^2}\Big[12\alpha(A - 3B) + 4e^{2Ar^2}\{-3 + 2\pi r^2 (4B_g + 3\gamma)\} -3e^{Ar^2}\{-4 + (A - 3B)(4\alpha + r^2)\}\Big]}{8 \pi r^2 (-1 + 3\beta)},\label{ro}
\end{eqnarray}
\begin{eqnarray}
p_r &=& \frac{e^{-2Ar^2}\Big[12\alpha(A - 3B)\beta + 4e^{2Ar^2}\{-3\beta + 2\pi r^2 (4\beta B_g + \gamma)\} -3\beta e^{Ar^2}\{-4 + (A - 3B)(4\alpha + r^2)\}\Big]}{8 \pi r^2 (-1 + 3\beta)}, \label{pr}
\end{eqnarray}
\begin{eqnarray}\label{pt}
p_t &=& \frac{e^{-2Ar^2}}{8\pi r^2 (-1 + 3\beta)}\Bigg[2e^{Ar^2}\{-1 - 3\beta + 4 \pi r^2 (4\beta B_g + \gamma) \} + 4\alpha[3A\beta + B\{-2 -3\beta + (3A -B)(-1 + 3\beta)r^2\}] +\nonumber\\ && e^{Ar^2}\Big(2 + 6\beta + \{A(2 - 9\beta) + 9B\beta\}r^2 + B(-A + B)(-1 + 3\beta)r^4 + 4\alpha\Big[-3A\beta + B\{2 + 3\beta +\nonumber\\ && (-A + B)(-1 + 3\beta)\}r^2\Big]\Big)\Bigg] 
\end{eqnarray}
Consequently, the components related to the SQM are as follows:
\begin{eqnarray}\label{rhoq}
\rho_q &=& \frac{e^{-2Ar^2}\Big[36\alpha(B - A\beta) + e^{2Ar^2}\{9 + 9\beta - 8\pi r^2 (4B_g + 3\gamma) \} - 9e^{Ar^2}\{1 + \beta + (B - A\beta)(4\alpha + r^2)\} \Big]}{8\pi r^2 (-1 + 3\beta)},
\end{eqnarray}

\begin{eqnarray}\label{pq}
p_q &=& \frac{e^{-2Ar^2}\Big[12\alpha(B - A\beta) + e^{2Ar^2}\{3 + 3\beta - 8\pi r^2 (4\beta B_g + \gamma) \} -3e^{Ar^2}\{1 + \beta(B- A\beta)(4\alpha + r^2)\}\Big]}{8 \pi r^2(-1 + 3\beta)}.
\end{eqnarray}

Now according to the junction conditions implying the continuity of the metric coefficients $g_{rr}$, $g_{tt}$ and $\frac{\partial g_{tt}}{\partial r}$ across the boundary surface $r={R}$ between the interior and the exterior regions, we find out
\begin{eqnarray}\label{grr}
e^{-AR^2} = \mathcal{F}({\mathfrak{R}}), 
\end{eqnarray}
\begin{eqnarray}\label{gtt}
 e^{B{\mathfrak{R}}^2 + C} = \mathcal{F}({\mathfrak{R}}),
\end{eqnarray}
\begin{eqnarray}\label{dgtt}
2B{\mathfrak{R}}e^{B{\mathfrak{R}}^2 + C} &=& \mathcal{F}^{'}({\mathfrak{R}}) \nonumber\\
&=& \frac{(-1 + \mathcal{W})\mathfrak{R}}{2\alpha \mathcal{W}}, 
\end{eqnarray}
where 
\begin{eqnarray}
\mathcal{W} = \sqrt{1 + \frac{8\alpha \mathcal{M}}{\mathfrak{R}^4}}.
\end{eqnarray}
Hence, solving equation (\ref{grr}), we get the value of $b$ as a function of the free model parameters $a$ and $\alpha$ as given below
\begin{eqnarray}\label{b}
  A &=& -\frac{1}{\mathfrak{R}^2}ln[K + \frac{\mathfrak{R}^2}{4\alpha}(1 - \mathcal{W})]
\end{eqnarray}

From Equations (\ref{gtt}) and (\ref{dgtt}) we get,
\begin{eqnarray}\label{Bvalue}
B = \frac{-1 + \mathcal{W}}{\mathcal{W}\{4\alpha K + (1 - \mathcal{W})\mathfrak{R}^2\}}
\end{eqnarray}
Furthermore, applying the condition $p_r (r = \mathfrak{R}) = 0$, we can write the parameters $\gamma$ in terms of the constants $A$, $B$, $\alpha$, $\beta$ and $B_g$:
\begin{eqnarray}\label{gamma}
\gamma &=& \frac{\beta e^{-2 A \mathfrak{R}^2} [-12 \alpha (A - 3 B) - 
   4 e^{2 A \mathfrak{R}^2} (-3 + 8 B_g \pi \mathfrak{R}^2) + 
   3 e^{A \mathfrak{R}^2} \{-4 + (A - 3 B) (4 \alpha + \mathfrak{R}^2)\}]}{8\pi \mathfrak{R}^2}.
\end{eqnarray}

Additionally, inserting the definition of $B$ in equation (\ref{gtt}) leads us to the following expression of the parameter $C$:
\begin{eqnarray}
    C &=& \rm{ln}[\mathcal{F}({\mathfrak{R}})] - B\mathfrak{R}^2 \label{Cval}.
\end{eqnarray}

\section{Celestial attributes}\label{Sec4}

To have a better understanding of the physical characteristics of the built celestial object, let us first establish some suitable values for the auxiliary model parameters.
In this work, to find the suitable values of the free parameters $A, B, C$, and $\gamma$, we take into consideration the compact star candidate 4U 1538 - 52.
 The mass and radius values of the strange spherical object 4U 1538 - 52 are $\mathcal{M} = 0.87 \pm 0.07~\mathcal{M}_{\odot}$ and $\mathfrak{R} = 7.866_{-0.21}^{+0.21}$ km \cite{rawls2011refined}. In order to construct a realistic stellar model with physical validity, we fix the values of the model parameters $\beta$, $K$, and $B_g$ as $\beta = 0.258$, $K= 0.951$ and $B_g = 70 \rm{MeV}/\rm{fm^3}$. In conclusion, we arrive at the numerical findings shown in Table \ref{table1} by using the newly discovered data of the 4U 1538 - 52 together with various values of the coupling parameter $\alpha \in [80, 120]$.

\begin{table}[h]
\centering
\caption{\label{table1} Some reasonable values of the free model parameters (Numerically calculated).}
\begin{tabular}{|c|c|c|c|c|}
\hline
$\alpha$ & $A~(\text{km}^{-2})$  & $B~(\text{km}^{-2})$ & $C$ & $\gamma$ \\
\hline
\hline
 80      & 0.00117579 & 0.000320026  & -0.0910055 & 0.0000526096    \\
 90      & 0.00117371 & 0.00031455   & -0.0905456 & 0.0000532271    \\
 100     & 0.00117167 & 0.000309277  & -0.0901009 & 0.0000538627   \\
 110     & 0.00116968 & 0.000304193  & -0.0896704 & 0.0000545154    \\
 120     & 0.00116773 & 0.000299289  & -0.0892536 & 0.0000551842   \\
\hline
\end{tabular}
\end{table}

\subsection{Regularity of the metric potentials}\label{mp}
To find a reliable and realistic model, metric potentials inside the hybrid star must be free of singularities. It is simple to verify that the metric potential values at the celestial object's center are obtained as\\
$\lim_{r\rightarrow 0}e^{\nu}=e^{C/2}$, a non-zero constant\\
$\qquad \lim_{r\rightarrow 0}e^{\lambda}=1.$  \\ 
These results lead to the metric potentials having finite values in the fluid configuration along with being singularity-free at the center of the structure.
In addition to this, at the center of the star, we have\\ 
$\frac{de^{\nu}}{dr}\Big\rvert_{r=0} = Br e^{(Br^2 +C)/2}\Big\rvert_{r=0} =0$\\
and
$\frac{d e^{\lambda}}{dr}\Big\rvert_{r=0} = Ar e^{Ar^2/2}\Big\rvert_{r=0} =0.$

Inside the star, the derivatives of the metric potential components are also positive and consistent.
The radial profiles of the metric coefficients, illustrated in Fig.~\ref{metric} make this simple to verify. 

\begin{figure}[H]
    \centering
        \includegraphics[scale=.63]{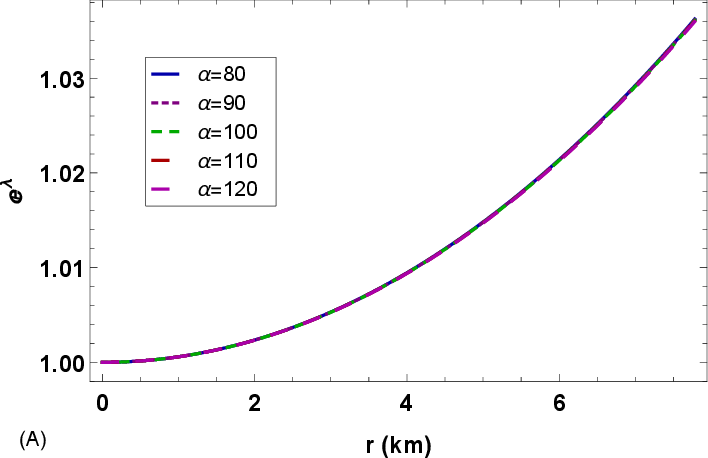}
         \includegraphics[scale=.63]{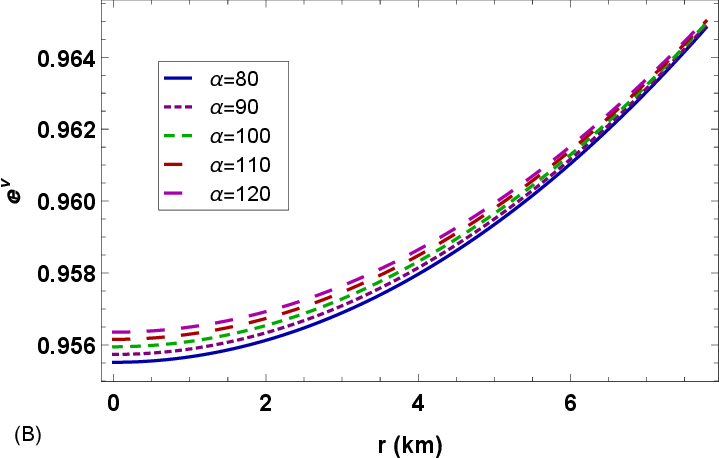}
        \caption{Variation of metric functions $e^{\lambda(r)}$ and $e^{\nu(r)}$ with respect to $r$.}\label{metric}
\end{figure}

Moreover,
$\frac{d^2 e^{\nu}}{d^2 r} \Big\rvert_{r=0} = B e^{C/2} >0,$\\
and  
$\frac{d^2 e^{\lambda}}{d^2 r} \Big\rvert_{r=0} = A >0,$\\
i.e. $e^{\nu}$ and $e^{\lambda}$ take their minimum values near the center and they gradually increase with $r$. Due to the non-singularity of the metric components under consideration at their center, we can validate that the metric coefficients behave properly in the range $(0, \mathfrak{R})$.

\subsection{Regularity of the fluid components}
The evolutional change of very dense strange star configurations is significantly influenced by physical variables like energy density $\rho$ and anisotropic stresses, i.e., $p_r$ and $p_t$. In order to demonstrate their viability at the center, these state determinants require finite and non-singular values. Also at the boundary of the
star radial pressure should vanish, i.e., $p_r (r = \mathfrak{R}) = 0$. In this connection, the central energy density and pressure are obtained respectively as
\begin{equation}\label{rhoc} 
\rho\Big\rvert_{r=0} = \rho_c = \frac{-3A \{5 + 4\alpha (A - 3 B)\} + 9B + 
 8\pi (4B_g + 3\gamma)}{8\pi (-1 + 3 \beta)},
\end{equation}
\begin{equation}\label{pc} 
 p_c = p_r\Big\rvert_{r=0} = p_t\Big\rvert_{r=0} =\frac{3\{3B + A(-5 -4A\alpha + 12\alpha B)\}\beta + 8\pi(4\beta B_g + \gamma)}{8\pi(-1 + 3\beta)}.
\end{equation}
The following relation can be used to calculate the surface energy density in addition to the results above.
\begin{eqnarray}\label{rhos} 
\rho\Big\rvert_{r=\mathfrak{R}} =\rho_s = \frac{e^{-2A\mathfrak{R}^2}[12\alpha(A-3B) + 4e^{2A\mathfrak{R}^2}\{-3 + 2\pi \mathfrak{R}^2(4B_g + 3\gamma)\} - 3e^{A\mathfrak{R}^2}\{-4 + (A - 3B)(4\alpha + \mathfrak{R}^2)\}]}{8\pi \mathfrak{R}^2(-1 + 3\beta)}.   \label{ro}
\end{eqnarray}

\begin{figure}[H]
    \centering
        \includegraphics[scale=.63]{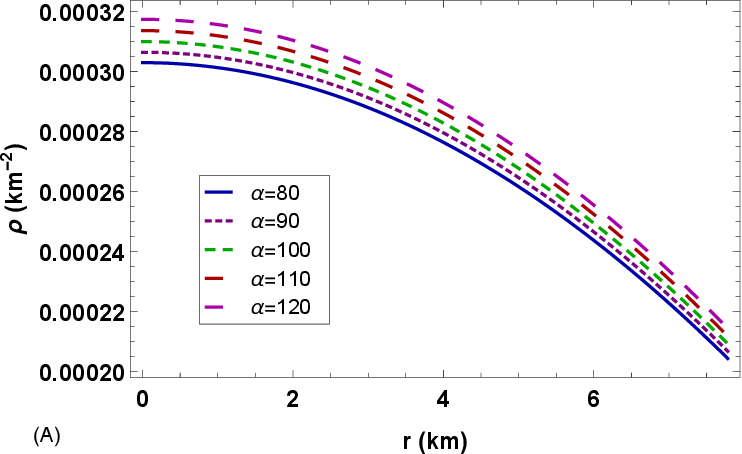}
         \includegraphics[scale=.63]{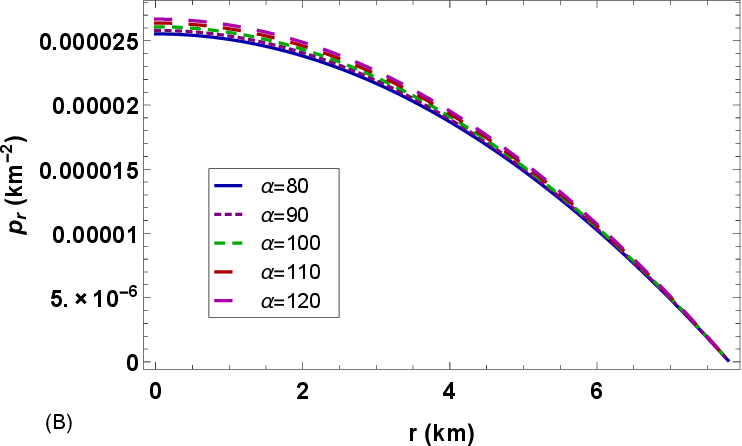}
         \includegraphics[scale=.63]{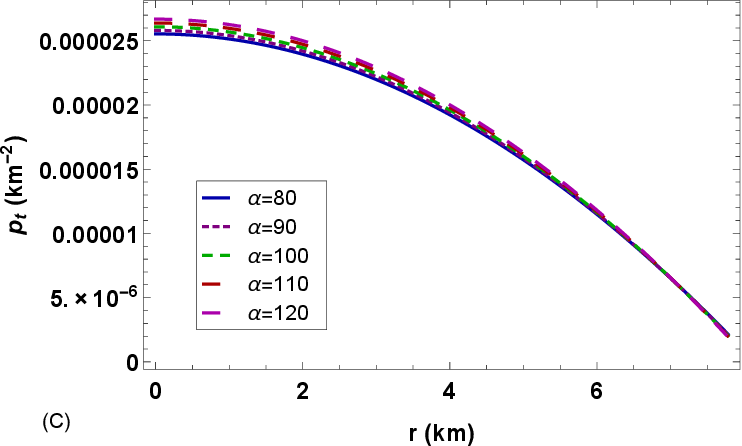}       
        \caption{Profiles of (A) matter density and (B, C) pressure components with respect to $r$.}\label{rho}
\end{figure}

Fig.~\ref{rho} displays the density and pressure profiles for various values of $\alpha$ and it indicates that they are all monotonically decreasing functions of `$r$' i.e. at the star's core, where they are at their highest value, and the boundary, where the radial pressure $p_r$ vanishes. At $r = \mathfrak{R}$, transverse pressure and density both exhibit positive values.
The central density grows as the coupling parameter $\alpha$ increases, which is another notable fact.

On the other side, Fig.~\ref{rhoq} displays the pressure and density profiles caused by quark matter. Here it is a very interesting fact that the quark matter density increases monotonically with radius `$r$'. In this connection, it is to be mentioned that under the consideration of non-interacting and interacting two-fluid models, pertaining to quark and baryonic matter, and between two extremes i.e. $p_q < 0$ and $p_q > 0$, there are $p_q$ values where gravity is attractive in the case of non-interacting objects but not in the case of interacting objects. Gravity in the halo is repulsive if $p_q < 0$, with a large enough absolute value, according to the analysis of Rahaman et al. \cite{rahaman2012quark}. For our model, we notice that pressure $p_q$ due to quark matter takes its negative value inside the fluid sphere. Bhar \cite{bhar2015new} and Abbas et al.\cite{abbas2021hybrid} also looked into the similar characteristics of quark matter density $\rho_q$. 

\begin{figure}[H]
    \centering
        \includegraphics[scale=.63]{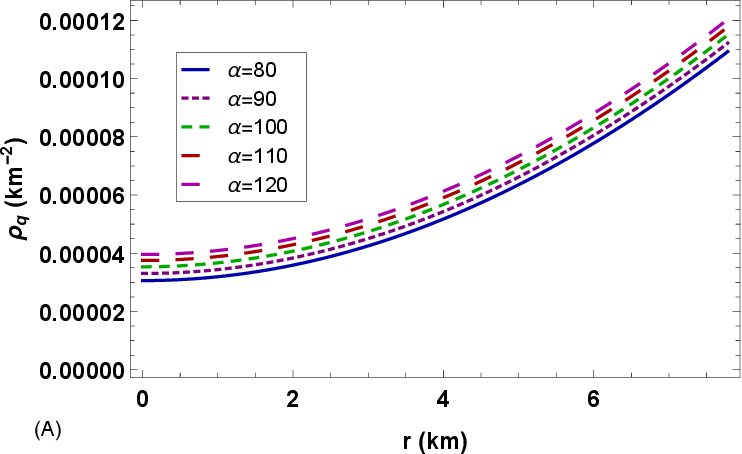}
         \includegraphics[scale=.63]{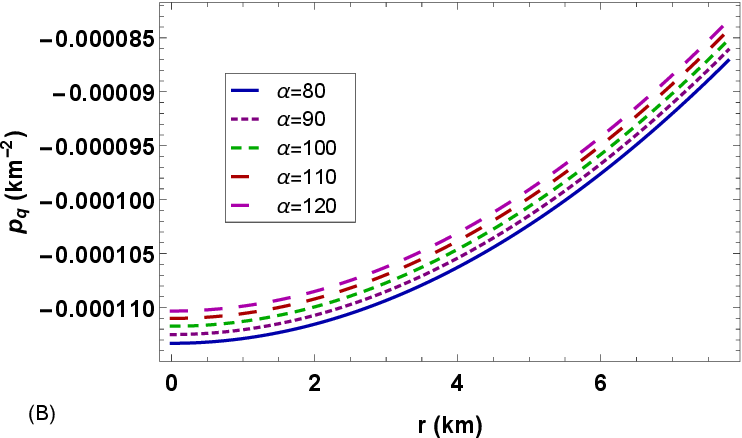}
        \caption{Profiles of (A) quark matter density and (B) quark matter pressure components with respect to $r$.}\label{rhoq}
\end{figure}

We provide some relevant numerical values for the central radial pressure $p_c$ and the central energy density $\rho_c$ in Table \ref{table2}. As a result, we can see that at the center, both the matter density and the tangential pressure take positive values. 

\begin{table}[h]
\centering
\caption{\label{table2} The numerical values of the central fluid components as well as the surface energy density.}
\begin{tabular}{|c|c|c|c|c|}
\hline
$\alpha$ & $\rho_c \times 10^{14}$ ($\text{gm}~\text{cm}^{-3}$) & $\rho_s\times 10^{14}$ ($\text{gm}~\text{cm}^{-3}$) & $p_c\times 10^{34}$ ($\text{dyne}~\text{cm}^{-2}$) \\
\hline
\hline
 80    & 3.02900 & 2.03913   & 3.09180276 \\
 90    & 3.06350 & 2.06306   & 3.12480478 \\
 100   & 3.09909 & 2.08770   & 3.15900534 \\
 110   & 3.13569 & 2.11300   & 3.19431968 \\
 120   & 3.17325 & 2.13892   & 3.23065096 \\
\hline
\end{tabular}
\end{table}

\subsection{Nature of the fluid components}
\begin{figure}[H]
    \centering
        \includegraphics[scale=.63]{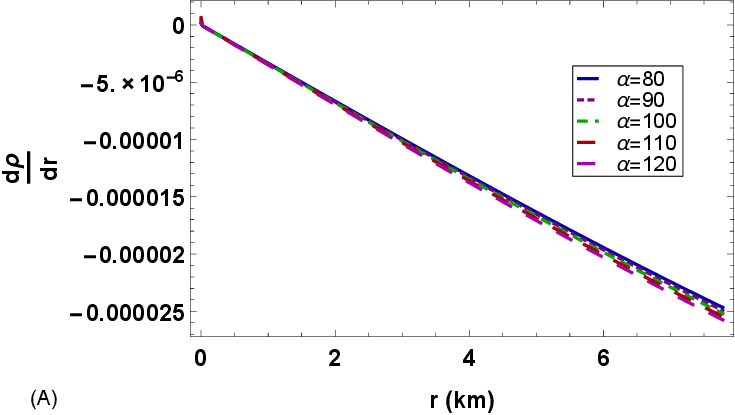}
         \includegraphics[scale=.63]{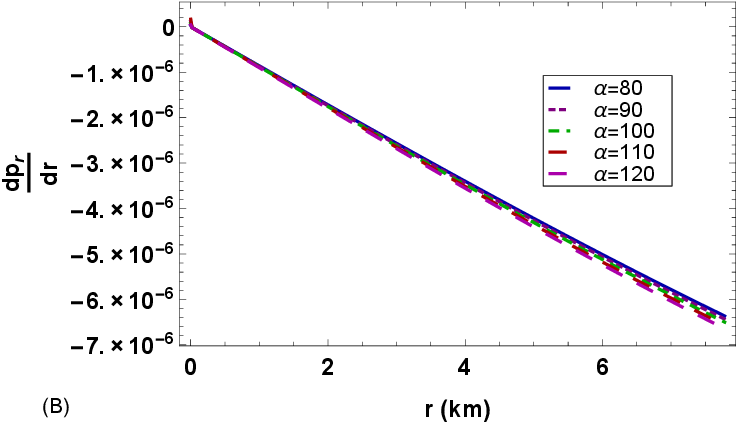}
         \includegraphics[scale=.63]{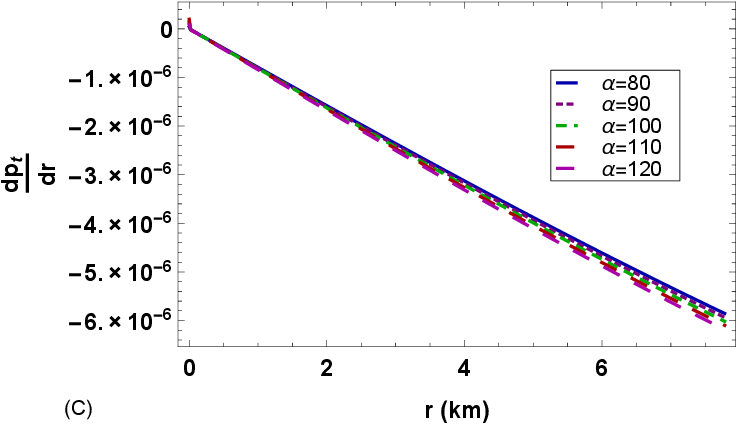}
         \caption{Gradients of the matter density and pressure components with respect to $r$.}\label{grad}
\end{figure}

We keep analyzing some actual elements of the stellar configurations by using the graphical analyses of density and pressure gradients, namely, $\frac{d\rho}{dr}, \frac{dp_r}{dr},$ and $\frac{dp_t}{dr}$, in order to test the validity of our model, which depicts a charged anisotropic compact celestial structure. We look at the evolution of the energy density and pressure gradients in Fig. \ref{grad} and come to the conclusion that their consistent negative behaviors inside the stellar object ensure the existence of an intensive celestial structure. 
We also highlight the fact that as the parameter $\alpha$ increases, the energy density and pressure gradients diminish.

\begin{figure}[H]
    \centering
         \includegraphics[scale=.63]{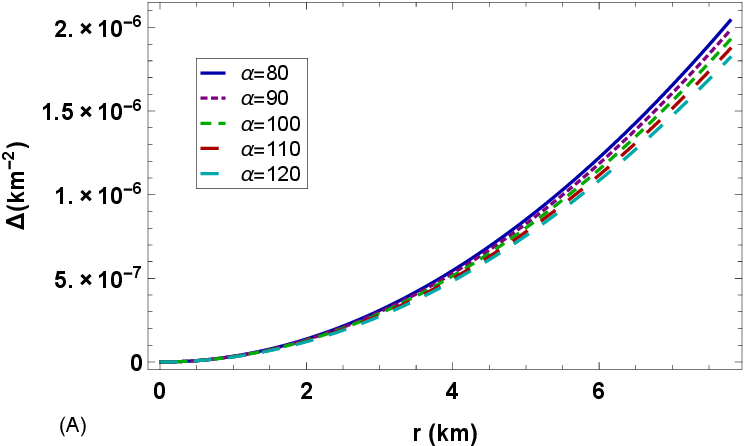}
         \includegraphics[scale=.63]{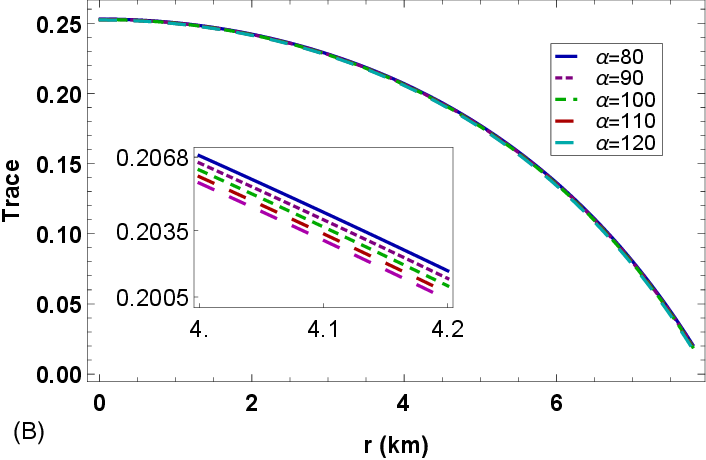}
        \caption{(A) Anisotropic factor ($\Delta$) and (B) the trace profile with respect to $r$.}\label{delta}
\end{figure}

Now we examine the effects of pressure anisotropy i.e. $\Delta = p_t - p_r$, on the compact celestial object described by the equation:
\begin{eqnarray}
\Delta &=& -\frac{3e^{-2Ar^2} \{-4\alpha Br - e^{2Ar^2} r + 
e^{Ar^2}(r + 4\alpha Br + Br^3)\}}{8\pi r^3} + \frac{1}{8\pi}\Bigg[\frac{e^{-2Ar^2}(-4\alpha B - e^{2Ar^2} + 12A\alpha Br^2 - 4\alpha B^2 r^2)}{r^2} + \nonumber\\&&  \frac{e^{-A r^2} (1 + 4\alpha B - 2Ar^2 + 3Br^2 - 4A\alpha Br^2 + 4\alpha B^2 r^2 - ABr^4 + B^2 r^4)}{r^2}\Bigg]    
\end{eqnarray}

The left panel of Fig. \ref{delta} depicts the type of pressure anisotropy for $ 80 \leq \alpha \leq 120 $ in our current model. When we look at the matter structure, the pressure anisotropy whose direction depends on the pressure components, becomes a fascinating critical factor. For $\Delta < 0$, i.e. $ p_t < p_r $, the anisotropy is directed inward or attractive. In contrast, when $\Delta > 0$, i.e. $ p_t > p_r $, the anisotropy is implied to be directed outward or repulsive \cite{hossein2012anisotropic1}. The characteristic of the pressure anisotropy is that it should disappear at the center of the star, indicating that the radial and transverse pressures are equal there; thus, the pressure turns isotropic at the star's core. Additionally, it must be positive within the stellar structure because a positive anisotropy produces a repulsive force that prevents the star from collapsing due to gravity \cite{gokhroo1994anisotropic}. It reaches its highest position from the center towards the boundary surface, and the value of $\Delta\Big\rvert_{r=\mathfrak{R}}$ is always positive for a physically plausible model because $p_t\Big\rvert_{r=\mathfrak{R}} > 0 $ and $p_r \Big\rvert_{r=\mathfrak{R}} = 0 $.\\
 
We exhibit the trace profile, or $\frac{p_r+2p_t}{\rho}$, of the celestial structure in the right panel of Fig. \ref{delta}. The most crucial thing to notice is that it decreases monotonically as one increases radial coordinates, with the highest values near the configuration's core, and remains positive throughout the fluid sphere. This finding supports the idea that stellar matter variables have a substantial profile, representing the compact surroundings of the possible stellar structure.
\cite{rej2023charged}.

\subsection{Mass-radius ratio and compactness}
\begin{figure}[H]
    \centering
\includegraphics[scale=.63]{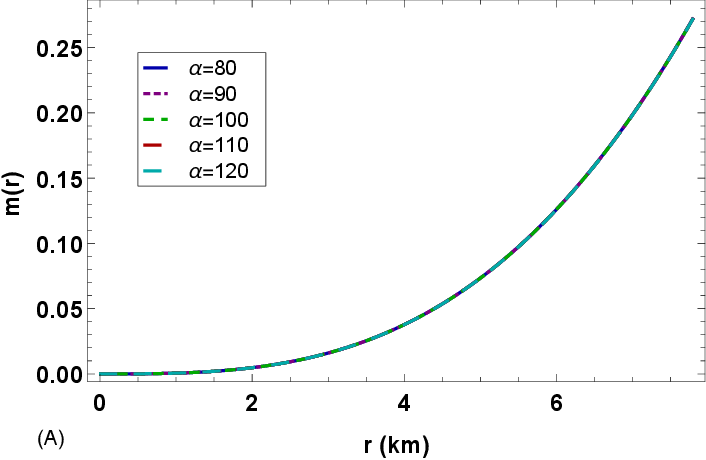}
\includegraphics[scale=.63]{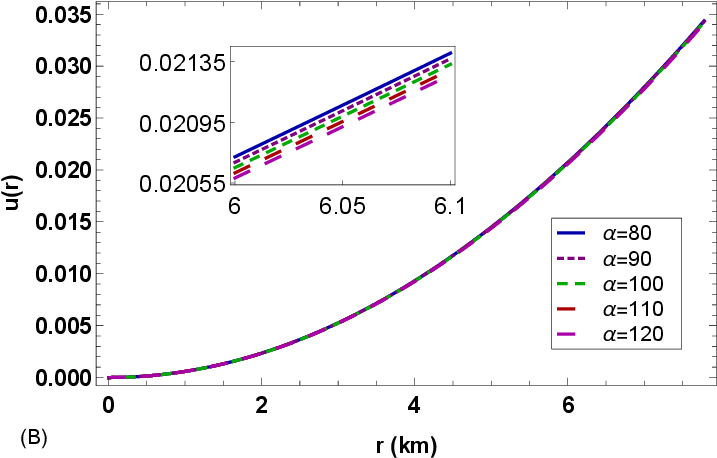}
\caption{(A) Mass function and (B) compactness factor against $r$.}\label{massu}
\end{figure}

Buchdahl demonstrated in \cite{Buchdahl1959general, rej2023charged} that the maximum permissible mass-radius ratio must comply with the inequality $\frac{2\mathcal{M}}{\mathfrak{R}} < \frac{8}{9}$ for a perfect fluid sphere devoid of any charge. In this case, $\frac{2\mathcal{M}}{\mathfrak{R}} = 0.329038 < \frac{8}{9}$ which leads again to a stable star model. 
By resolving the subsequent integral in the specified system, the mass function $m(r)$ can be found such as:

\begin{eqnarray}\label{mm}
 m(r)=4\pi\int^r_0{\rho \xi^2 d\xi} 
 \end{eqnarray}
The gravitational mass $\mathcal{M}$ at the boundary of our star is thus obtained using the aforementioned expression as follows:
\begin{eqnarray}\label{mm1} 
\mathcal{M}= m(r)\Big\rvert_{r=\mathfrak{R}}=\frac{\mathfrak{R}}{2}\Big[1-e^{-A\mathfrak{R}^2}\Big].
\end{eqnarray}
Also, the effective mass can be computed as
\begin{eqnarray}\label{mef}
\mathcal{M}^{eff} = \int^{\mathfrak{R}}_0{4\pi \rho \xi^2 d\xi} = \frac{\mathfrak{R}}{2}\Big(1-e^{-A\mathfrak{R}^2}\Big),
\end{eqnarray}
 However, the compactness factor $u(r)$ and its actual form are also significant items and have been described as:
\begin{eqnarray}\label{ur}
 u(r) = \frac{m(r)}{r},   
\end{eqnarray}
and
\begin{eqnarray}\label{ur2}
 u^{eff}=\frac{\mathcal{M}^{eff}}{\mathfrak{R}}.    
\end{eqnarray}
The behaviors of the mass function and the compactness factor are demonstrated in Fig. \ref{massu}. Both of the above quantities are evidently regular, positive, and monotonically increasing with the increase of `$r$' within the stellar medium. Moreover, $u(r)$ decreases when the coupling parameter $\alpha$ increases.
It is also to be noted that $u = 10^{-3}$ for white dwarfs and $u = 10^{-5}$ for an average star. Additionally, it has been predicted that $u = 0.5$ for a black hole, $u \in (0.1, 0.25)$ for a strange neutron star, and $u \in (0.25, 0.5)$ for an ultra-compact stellar object\cite{bhar2023physical}. 

\subsection{Redshift functions}
In terms of cosmology and astronomy, redshift is a highly significant phenomenon as it makes it easy to explore the characteristics of our galaxy and possibly the entire universe.
This phenomenon happens when the electromagnetic radiation of an object shifts towards the less energetic (higher wavelength) at the end of the spectrum. 
In this manner, the gravitational redshift, also known as inner redshift, more often referred to as a fall in wave frequency and a rise in wavelength, results from the loss of energy experienced by electromagnetic waves or photons leaving a gravitational field, especially at the surface of a massive star. At the same time, the opposite shift, referred to as a blueshift or negative redshift, is defined as a drop in wavelength and an increase in frequency and energy. The interior density profile is also explained by the behavior of inner redshift or more generally it is useful in assessing the nature of the strong relationship between the substance of a celestial structure and its EoS. A photon that exits the center and goes to the surface must travel much farther through the dense core area which leads to more dispersion and results in a loss of energy. Whereas a photon that comes out from near the surface will travel a shorter path through a denser region, resulting in less dispersion and less energy loss. As a result, the center and surface have the highest and lowest inner redshifts, respectively. Nevertheless, the total mass and radius-that is, the surface gravity determine the surface redshift. Generally, it connects a compact star's mass and radius. It is a very interesting fact that when mass increases, the radius will also increase slightly, which will yield more surface gravity and more surface redshift. Thus, the trends of inner and surface redshifts are opposite \cite{rej2023charged}.
Inner redshift $Z_g$  is defined by the expression
\begin{eqnarray}
   Z_g(r) = e^{-\nu(r)} - 1 = e^{-\frac{Ar^2}{2}} - 1 
\end{eqnarray}
and with the help of equation (\ref{ur}), the surface redshift function $Z_s$ is represented by the formula
\begin{eqnarray}
    Z_s = \frac{1}{\sqrt{1 - u(r)}} - 1.
\end{eqnarray}

\begin{figure}[H]
    \centering
\includegraphics[scale=.63]{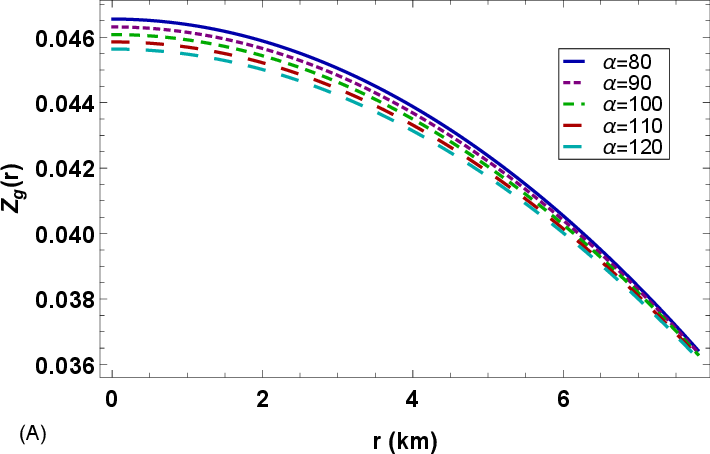}
\includegraphics[scale=.63]{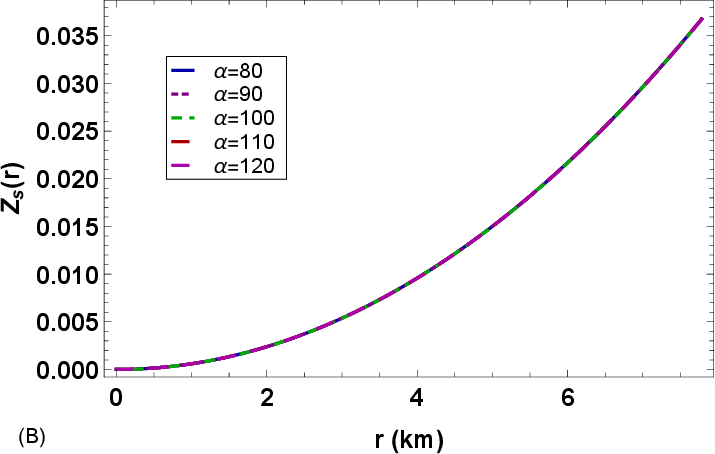}
\caption{Behavior of the (A) inner and (B) surface redshift functions against $r$.}\label{redshift}
\end{figure}

The profiles of both redshifts have been shown in Fig.~\ref{redshift}. Clearly, they are regular positive, and surface redshift is a monotonically increasing function while the inner redshift decreases with `$r$'. With a restriction $Z_s \leq 5.211$, the maximum surface redshift of an anisotropic fluid sphere is predicted to occur\cite{bohmer2006bounds}. The right panel of Fig. \ref{redshift} indicates the surface redshift is valid for our stellar object and that it satisfies the criteria throughout the stellar configuration.

In Table~\ref{table3}, for different values of the coupling parameter $\alpha$, we give a comparative numerical analysis of the effective compactness ($u^{eff}(\mathfrak{R})$) and surface redshift function($Z_s(\mathfrak{R})$).

\begin{table}[h]
\begin{center}
\caption{\label{table3} Comparative study of effective compactness and surface redshift.}
\begin{tabular}{|c|c|c|c|}
\hline
$\alpha$   &   $u^{eff}(\mathfrak{R})$  &  $Z_s(\mathfrak{R})$ \\
\hline\hline
80  & 0.0345182 & 0.0364149 \\
90  & 0.0344592 & 0.0363492 \\
100 & 0.0344015 & 0.0362850 \\
110 & 0.0343451 & 0.0362223 \\
120 & 0.0342899 & 0.0361609 \\
\hline
\end{tabular}
\end{center}
\end{table}

\subsection{EoS parameter: Zeldovich's condition}

\begin{figure}[H]
    \centering
\includegraphics[scale=.63]{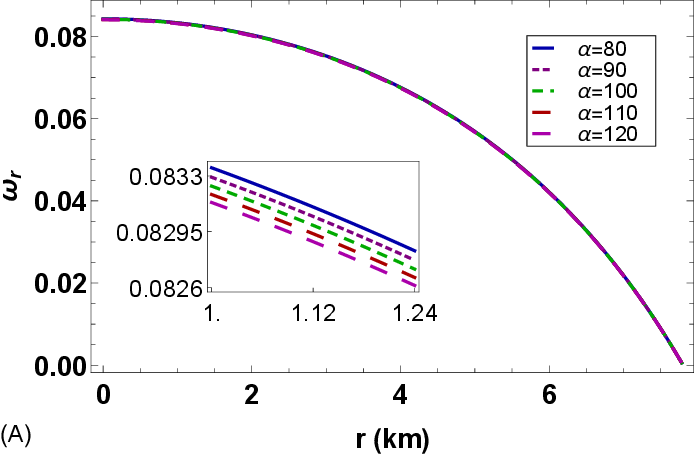}
\includegraphics[scale=.63]{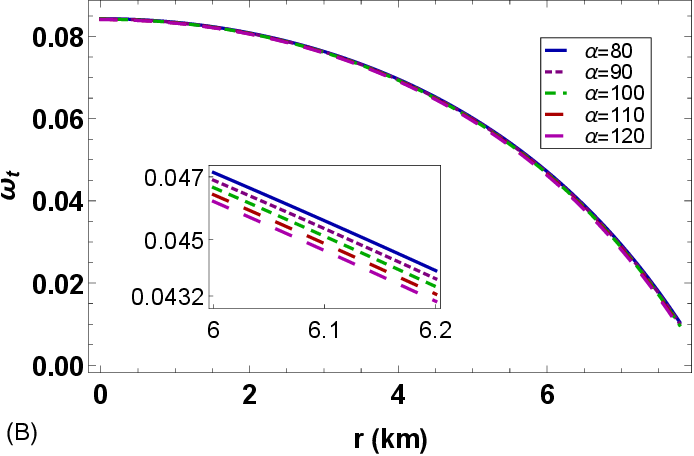}
\includegraphics[scale=.63]{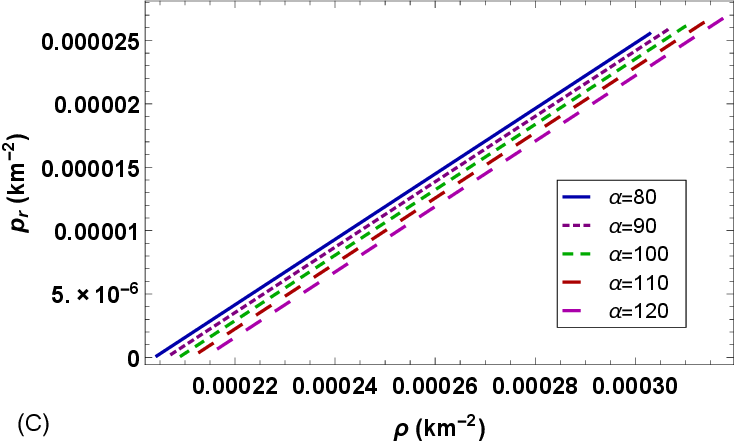}
\includegraphics[scale=.63]{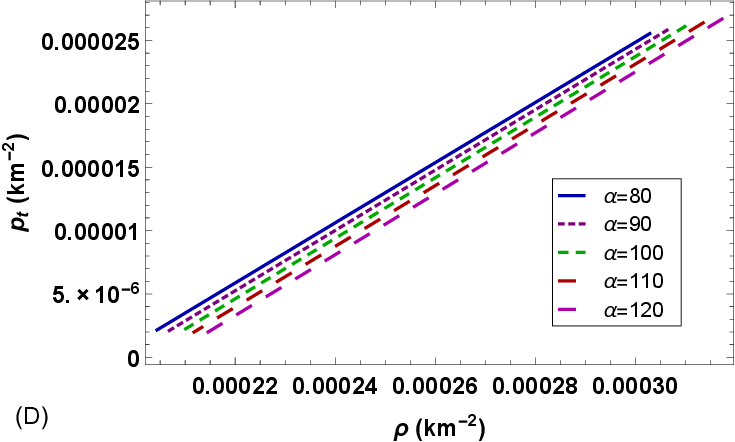}
\caption{Evolution of $\omega_r$ and $\omega_t$ with respect to $r$ as well as behavior of $p_r$ and $p_t$ with respect to $\rho$.}\label{eos}
\end{figure}

The question of the equation of state of matter at ultrahigh densities is currently the subject of greater discussion in relation to the problem of the gravitational collapse of heavy star evolution in its last stage. The effectively combined $\omega (= p/\rho)$ of the entire amount of matter in the universe is determined by weighing the contributions of various elements, including the dark and baryonic slow-moving matter $(\omega =0)$, ultra-relativistic matter $(\omega =1/3)$, and dark energy (probably $\omega \sim -1)$. Consequently, reliable measurements of $\omega$ will also hint at the relative diversity of various materials.
As to Zeldovich's criterion, any physically acceptable fluid sphere must have a pressure-to-density ratio that is positive and smaller than unity, meaning that $\omega$ should lie between 0 and 1 everywhere and be continuous at the junction \cite{l1962equation, zeldovich1971relativistic}.
So, two dimensionless quantities that can be used to describe the relationship between matter density and pressure are referred to as the EoS parameters, usually written as $$\omega_r = \frac{p_r}{\rho},\omega_t = \frac{p_t}{\rho}.$$
By resolving the field equations, it is accepted that the radial pressure and matter density are linearly connected in the model, but the transverse pressure and matter density are not yet known.
We have depicted their profiles in Fig. \ref{eos} in order to examine the behavior of the equation of state parameter.
The graphs demonstrate that $\omega_r$ and $\omega_t$ both are monotonically decreasing functions of `$r$', but they both fall inside the range $0 < \omega_r, \omega_t < 1.$ 
This observation leads to the significant conclusion that energy density diminishes faster than the volume expands in an expanding cosmos, due to the wavelength of the radiation being red-shifted. Regarding the fact that we are working with a hybrid star that contains both baryonic and strange quark matter, it is found that the significant EoS parameter components have positive values. Additionally, the graphic shows how the radial and transverse pressures have changed in response to density.
So, it is simple to show that our model of the celestial objects accurately portrays the realistic character of the composition of matter under the radial and transverse components of the EoS. 
Finally, we can easily check this fact from the sub-figures (A) and (B) of Fig.~\ref{eos}, which confirms the stability of our proposed model.

\section{Stability and feasibility analysis}\label{Sec5}
An essential mechanism, the stability mechanism, is examined in this section.
For the model to be an attainable, regular, and realistic hybrid stellar model with a given radius, it must meet the stability conditions listed below inside the stellar configuration. Maintaining the consistency of the model may be difficult because several variables may change at once.
In this respect, novel techniques for assessing the stability of a stellar model include Herrera's cracking concept (or the causality condition), relativistic adiabatic index, Energy Conditions (ECs), and static equilibrium through the TOV equation.

\subsection{Herrera's cracking concept}
The subliminal sound speed of the pressure waves must be less than the speed of light in any scenario explaining the stellar interior. When dealing with anisotropic fluids, the pressure waves propagate in the radial and transverse directions, which are the major axes of the object.
The speed of the subliminal sound in these directions is determined by 
$$V_r = \sqrt{\frac{dp_r}{d\rho}}, V_t = \sqrt{\frac{dp_t}{d\rho}}$$

In order to produce a physically realistic model, it is crucial to confirm the causality condition, which specifies that the speed of sound inside the compact object should be less than unity everywhere, i.e.
$0 \leq V_r^2 \leq 1 $ and $0 \leq V_t^2 \leq 1$ \cite{karmakar2023charged}. 
From equation (\ref{eos1}), we can instantly conclude that-
\begin{eqnarray}
{V_r}^2 = \beta
\end{eqnarray}
Also from the equation (\ref{pt}), we have
\begin{eqnarray}
 V_t^2 &=& \Big(\frac{dp_t}{dr}\Big)\Big/\Big(\frac{d\rho}{dr}\Big)= \frac{V_1}{V_2} 
\end{eqnarray}

\begin{figure}[H]
\centering
\includegraphics[scale=.63]{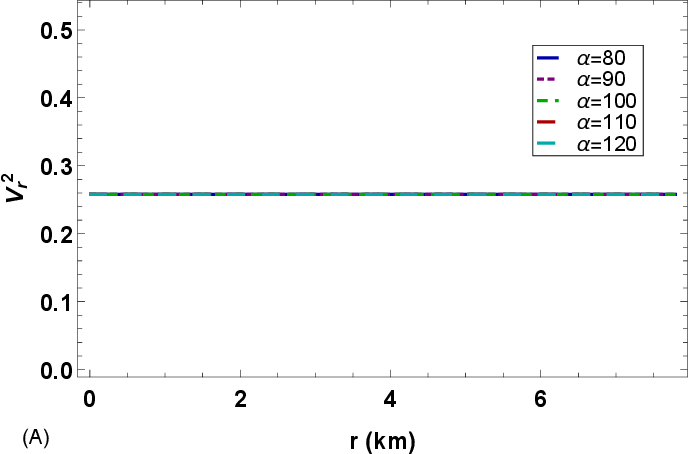}
\includegraphics[scale=.63]{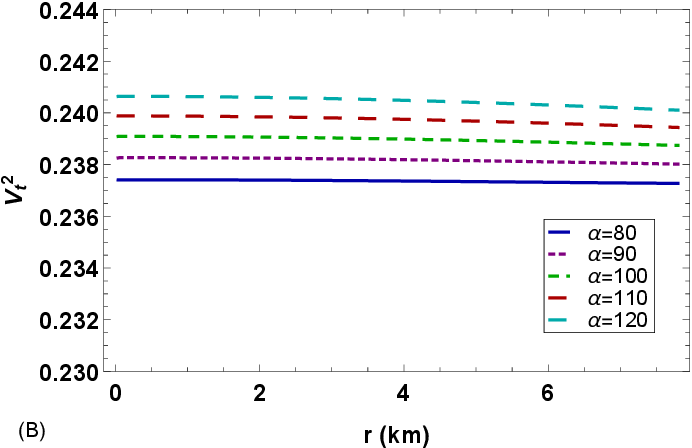}
\includegraphics[scale=.63]{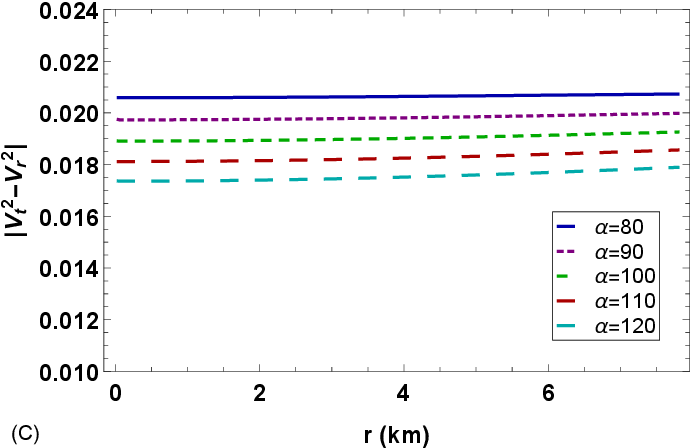}
\caption{(A, B) Square of the sound velocity components and (C) the stability factor $|V_t^2-V_r^2|$ have been plotted against radius $r$.}\label{speed}
\end{figure}

Where 
\begin{eqnarray*}
V_1 &=& 2(1 + 3 \beta) e^{2Ar^2} + 
   4\alpha [2B - 3A\beta + 3B\beta + 
      2A\{-3A\beta + B(2 + 3\beta)\}r^2 +  
      2AB(-3 A + B) (-1 + 3 \beta) r^4] + \nonumber\\&&
   e^{Ar^2}\Big(-2[1 + 3\beta + \alpha \{-6A\beta + B(4 + 6\beta)\}] + 
      2A[-1 - 3\beta + 
         \alpha \{6A\beta - 2B(2 + 3\beta)\}]r^2 + [B^2(-1 + 
            3\beta) \nonumber\\&& + AB\{1 + 4\alpha B - 12 (1 + \alpha B)\beta\} + 
         A^2\{-2 + 9\beta + 4\alpha B(-1 + 3\beta)\}]r^4 + 
      A(A - B)B(-1 + 3\beta)r^6\Big),
\end{eqnarray*}
and
\begin{eqnarray*}
V_2 &=&  3\Big[4e^{2Ar^2} - 4\alpha(A - 3B)(1 + 2Ar^2) + 
     e^{Ar^2} \{-4 + 4\alpha(A - 3B) + 
        4A(-1 + A\alpha - 3\alpha B)r^2 \nonumber\\&& + A(A - 3B) r^4\}\Big]. 
\end{eqnarray*}
As a result, while the tangential velocity varies across the star, the radial velocity remains constant.
In reality, the slope of the $p_r(\rho)$ and $p_t(\rho)$ functions is what determines the sound velocity. The upper left and right panels of Fig. \ref{speed} display the profiles of $V_r^2$ and $V_t^2$ for various values of $\alpha$.
These profiles clearly show that both sound velocities are within the expected range $[0,1]$. We may therefore claim that our model satisfies the causality criterion. 

In this connection, Herrera and his collaborators presented the idea of cracking for the self-gravitating by utilizing the ideal fluid and anisotropic material distributions in their publications \cite{herrera1992cracking, di1994tidal, di1997cracking}. Such a unique idea was developed to explain how the existence of non-vanishing radial forces causes fluid distributions to alter immediately as they leave their equilibrium phases. This technique is useful for detecting anisotropic matter configurations that might be unstable.
It is simple to explain this method as follows:
\begin{eqnarray}
\frac{\delta\Delta}{\delta\rho} \sim \frac{\delta (p_t - p_r)}{\delta\rho} \sim \Big(\frac{\delta p_t}{\delta\rho} - \frac{\delta p_r}{\delta\rho}\Big) \sim \big(V_t^2 - V_r^2\big)
\end{eqnarray} 
Consequently, 
\begin{eqnarray}\label{crack}
|V_t^2 - V_r^2| \leq 1
\end{eqnarray}
results from the causality conditions $0 \leq V_r^2 \leq 1$ and $0 \leq V_t^2 \leq 1$. 
Moreover, the equation(\ref{crack}) can be explicitly interpreted as \cite{jasim2021anisotropic}:
\begin{equation}\label{crack1}
|V_t^2 - V_r^2| \leq 1 =\left\{\begin{array}{ll}
                    -1 \leq V_t^2 - V_r^2 \leq 0,  & \qquad {\rm Potentially ~ stable}, \\
                   0 < V_t^2 - V_r^2 \leq 1, & \qquad {\rm Potentially ~ unstable}.
                  \end{array}
\right.
\end{equation}
Actually, the prospect of stability was investigated in the area of the star's interior where the radial velocity of sound is faster than the transverse velocity. As a result, from the lower panel of Fig. \ref{speed}, we can see that the radial speed of sound is faster than the tangential speed of sound throughout the star i.e. $|V_t^2 - V_r^2| \leq 1$ for $\alpha = 80, 90, 100, 110$ and $120$ in $[0, \mathfrak{R}]$ which demonstrates that the celestial core is not cracked \cite{andreasson2009sharp}.
As a result, it may be inferred from Fig. \ref{speed} that the present hybrid star model is theoretically stable because it complies with both the causality requirements and Herrera's cracking idea.

\subsection{Relativistic adiabatic index}
\begin{figure}[H]
\centering
\includegraphics[scale=.63]{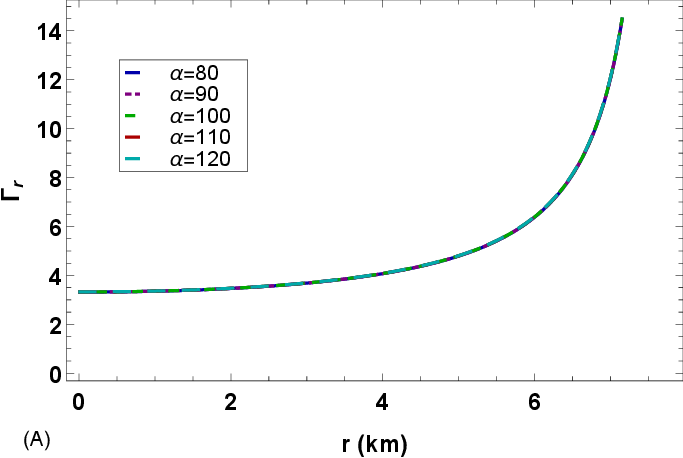}
\includegraphics[scale=.64]{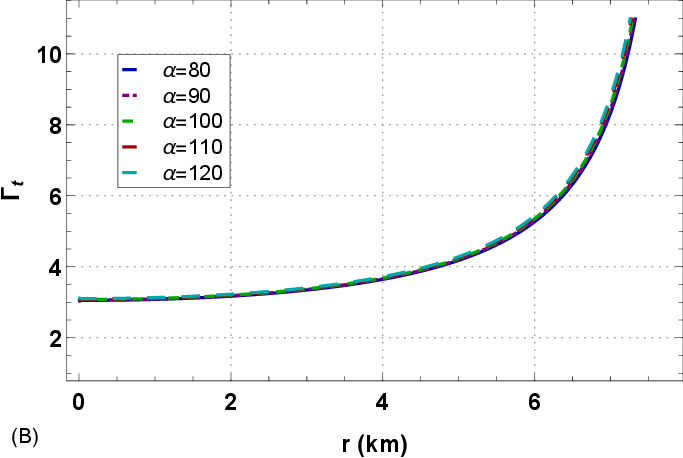}
\caption{Behavior of the adiabatic index $\Gamma_r$ and $\Gamma_t$ as a function of the radial distance $r$.}\label{gama}
\end{figure}

In order to study the region of stability of the hybrid star model, we will analyze a key and significant ratio of the two particular conditions provided by $\Gamma$. When pressure anisotropy exists, the expressions for the adiabatic index are as follows:
\begin{eqnarray}
\Gamma_r= \frac{\rho + p_r}{p_r}V_r^2,
\end{eqnarray} 
and 
\begin{eqnarray}
\Gamma_t= \frac{\rho + p_t}{p_t}V_t^2,
\end{eqnarray} 
Depending on the variational method, Chandrasekhar \cite{chandrasekhar1964dynamical} proposed a way to explore dynamical stability, and an essential connection for the adiabatic index was found using this methodology \cite{chandrasekhar1964dynamical, merafina1989systems}.
In this regard, several scholars have concentrated their efforts on examining the dynamical stability of the stellar formations \cite{Heint, Hilleb, Bombaci}. As a positive anisotropy factor may retard the development of an instability, the stability requirement for a relativistic compact object is defined by $\Gamma > 4/3$ in the presence of a positive and growing anisotropy factor $\Delta = p_t - p_r > 0$ \cite{heintzmann1975neutron}. lately, Moustakidis \cite{moustakidis2017stability} improved this idea to include radiation and anisotropy.
As seen in Fig. \ref{gama}, the behavior of $\Gamma$ shows that the adiabatic index is greater than $4/3$ inside the stellar interior for our proposed model, indicating that our model is stable from the relativistic adiabatic
index point of view. Also, we see that both $\Gamma_t$ and $\Gamma_r$ are monotonically increasing functions of `$r$'. 

For various coupling parameter $\alpha$ values, we provide a numerical comparison of the adiabatic index $\Gamma_r$ and $\Gamma_t$ at $r=0$ in Table \ref{table4}.
 
\begin{table}[h]
\begin{center}
\caption{\label{table4} Comparative study of the adiabatic index $\Gamma_r$ and $\Gamma_t$ at $r=0$.}
\begin{tabular}{|c|c|c|c|}
\hline
$\alpha$   &   $\Gamma_r$  &  $\Gamma_t$ \\
\hline\hline
80  & 3.318 & 3.05326 \\
90  & 3.32017 & 3.06626 \\
100 & 3.3222 & 3.07872 \\
110 & 3.32412 & 3.09069 \\
120 & 3.32595 & 3.10216 \\
\hline
\end{tabular}
\end{center}
\end{table}
\~~~~\

\subsection{Energy Conditions}
The energy conditions are a key component of this study since they must be fulfilled for each internal fluid sphere and be non-negative over the entire stellar medium. One can distinguish between a typical substance and an unusual fluid by applying the energy conditions (ECs) to substance parameters. Simply, The pressures and density need to be limited to some bound in order to appear physically suitable. 
For the current model to have any physical significance in this approach, the following energy constraints must be executed:

\begin{itemize}
\item Strong Energy Condition (SEC):~ $\rho + p_r + 2p_t  \geq 0$\\

\item Weak Energy Condition (WEC):~ $\rho + p_r \geq 0, ~~~ \rho + p_t  \geq 0$\\

\item Null Energy Condition (NEC):~ $\rho  \geq 0$ \\

\item Dominant Energy Condition (DEC):~ $\rho - p_r  \geq 0,~~~\rho - p_t \geq 0$ \\ 

\item Trace Energy Condition (TEC):~ $\rho - p_r  - 2p_t \geq 0$ \\ 
\end{itemize}

\begin{figure}[H]
\centering
\includegraphics[scale=.6]{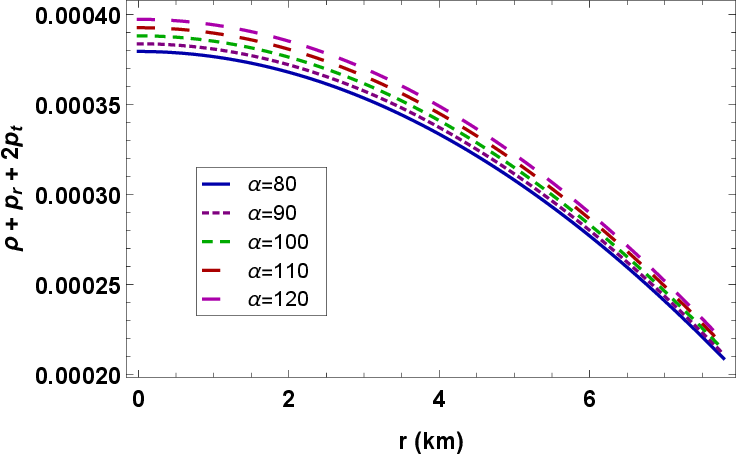}
\includegraphics[scale=.6]{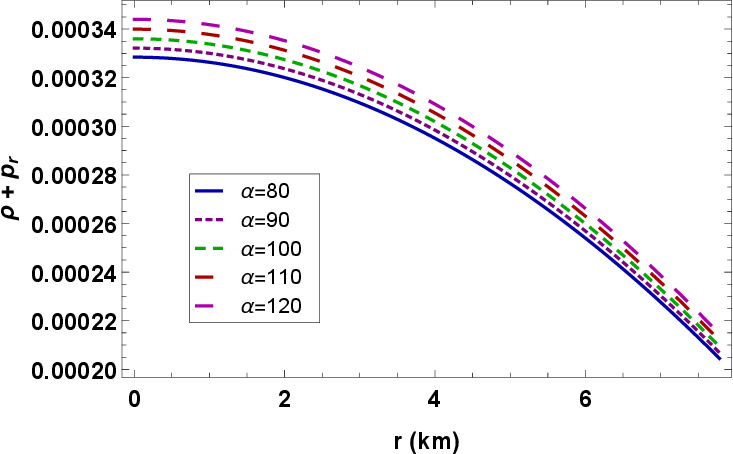}\\[1\baselineskip]
\includegraphics[scale=.6]{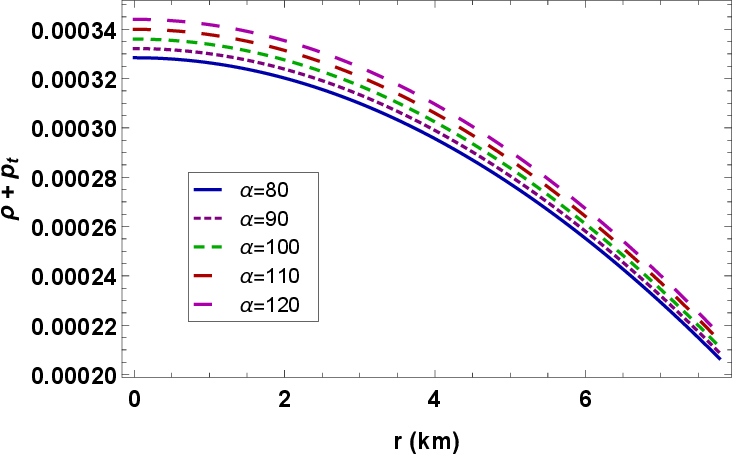}
\includegraphics[scale=.6]{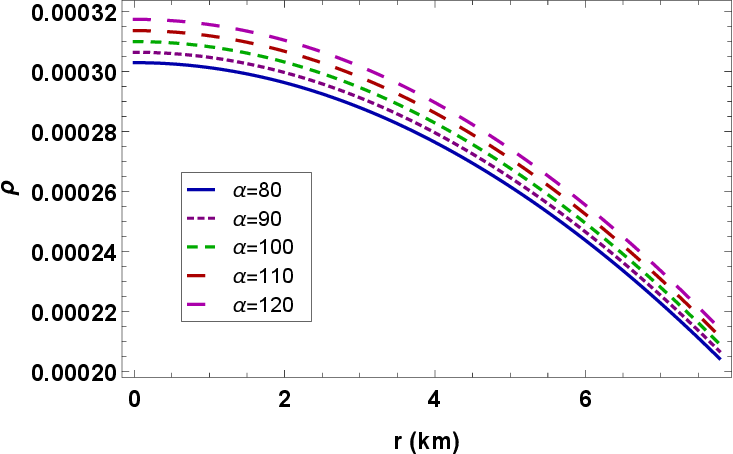}\\[1\baselineskip]
\includegraphics[scale=.6]{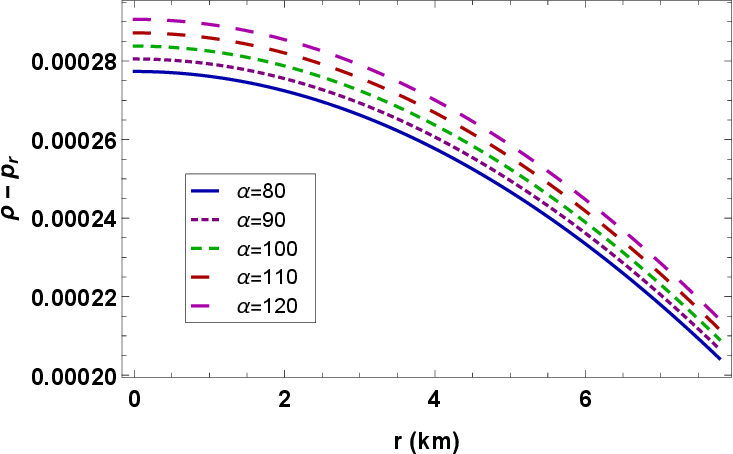}
\includegraphics[scale=.6]{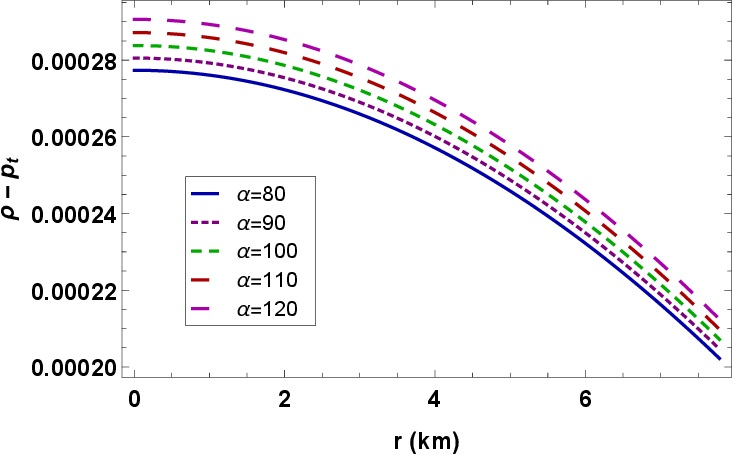}\\[1\baselineskip]
\includegraphics[scale=.6]{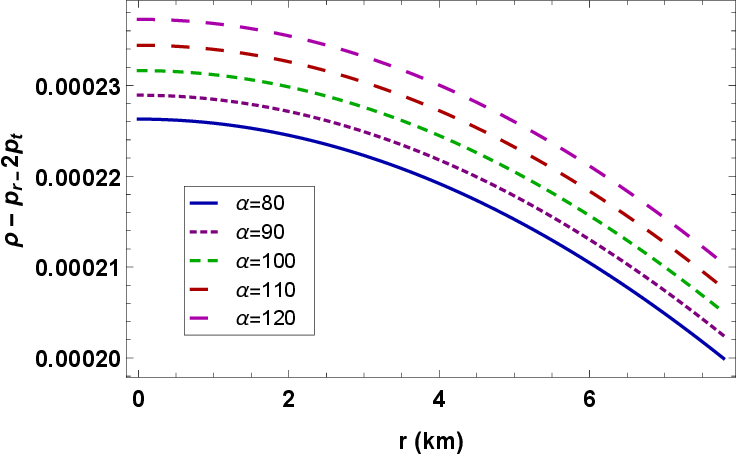}
\caption{Variation of the ECs with respect to the radial coordinate $r$.}\label{Ener}
\end{figure}

We can observe from their graphical representations in Fig. \ref{Ener} that the profiles of the aforementioned energy criteria are entirely satisfied, indicating that the suggested compact star model is physically stable.

\section{Hydrostatic equilibrium}\label{Sec6}
\begin{figure}[H]
\centering
\includegraphics[scale=.62]{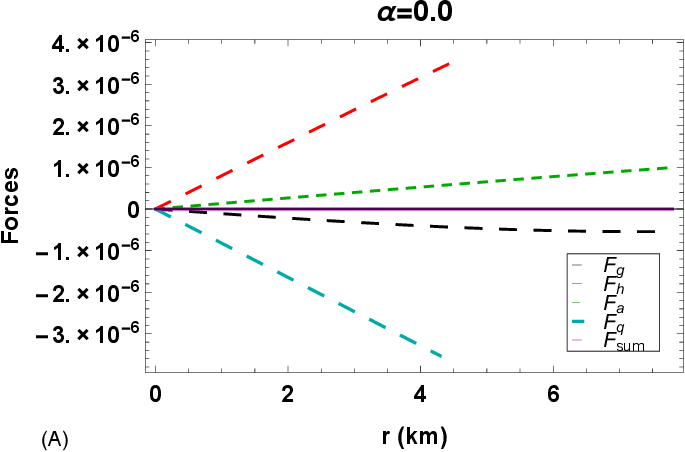}
\includegraphics[scale=.62]{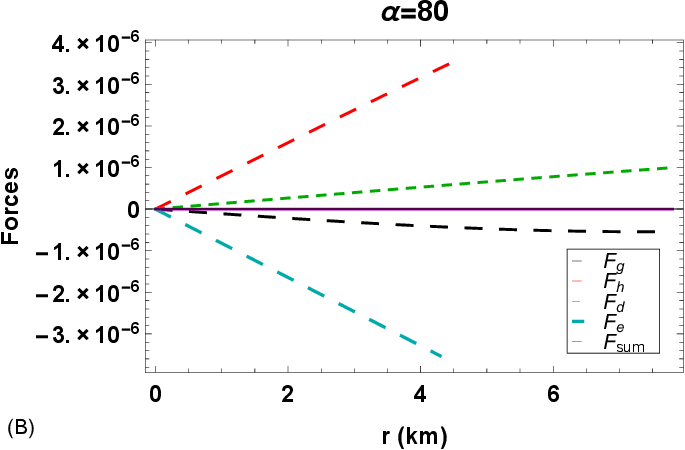}\\
\includegraphics[scale=.62]{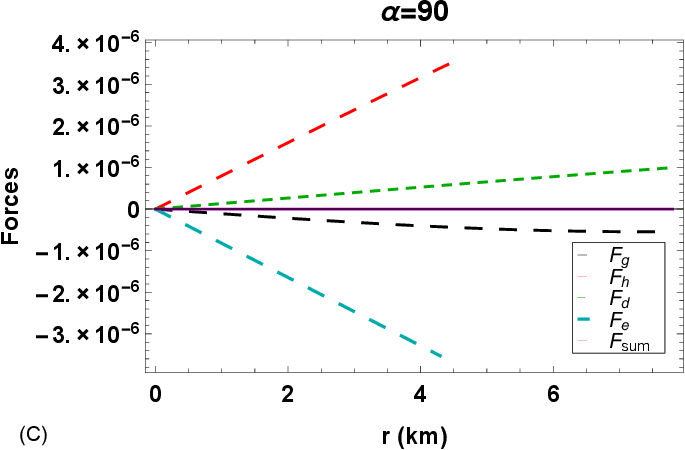}
\includegraphics[scale=.62]{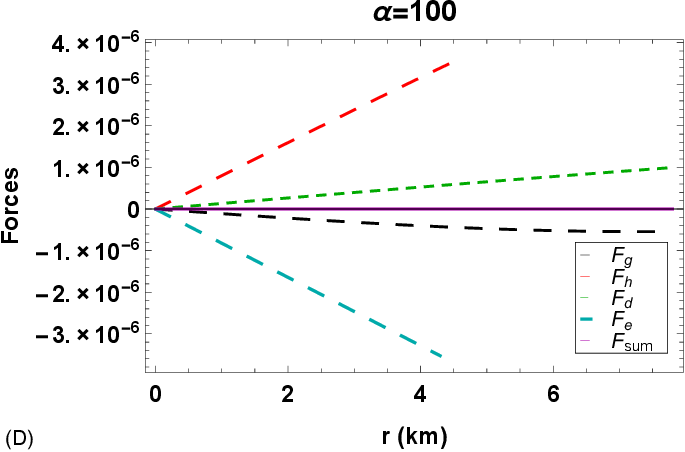}\\
\includegraphics[scale=.62]{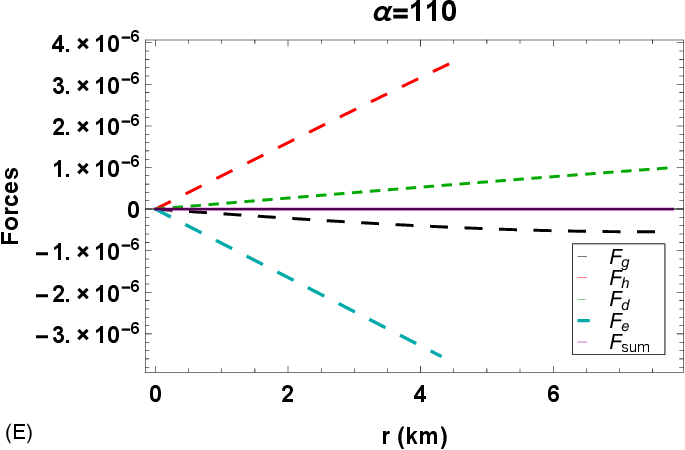}
\includegraphics[scale=.62]{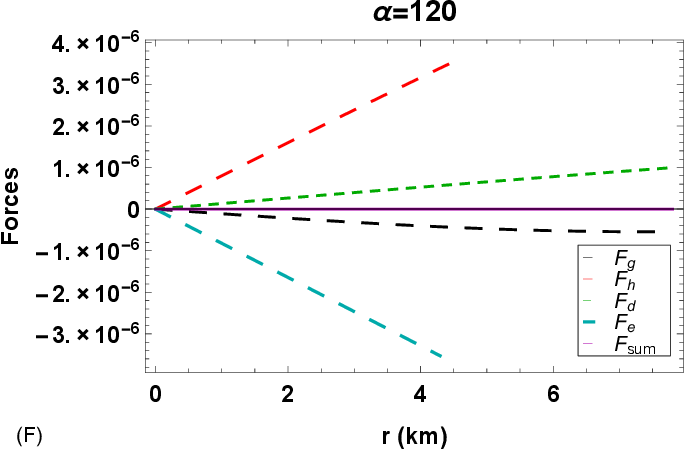}
\caption{Variations of forces with respect to $r$ for different values of $\alpha$.}\label{force}
\end{figure}

The hydrostatic equilibrium equation is a crucial characteristic of the physically realistic compact object supplied. To assess this state of equilibrium equation for our compact star candidate under the combined action of various forces, the generalized Tolman-Oppenheimer-Volkov (TOV) equation can be used followed by the equation  \cite{tolman1939static, oppenheimer1939massive}:

\begin{equation}\label{tov1}
F_{sum}=F_g + F_h + F_a + F_q=0,
\end{equation}
where $$F_g = -\frac{\nu'}{2}(\rho+p_r),$$ 
$$F_h = -\frac{dp_r}{dr},$$ 
$$F_a = \frac{2}{r}(p_t - p_r)$$ and
$$F_q=-\frac{\nu'}{2}(\rho_q + p_q) - \frac{dp_q}{dr}$$ 
respectively, correspond to the gravitational force, the hydrodynamic force of ordinary matter, the anisotropic force, and the force associated with quark matter. 
We provide the profile of forces $F_g$, $F_h$, $F_a$, and $F_q$ as well as $F_{sum}$ in Fig.\ref{force} for different values of $\alpha$ to secure the state of equilibrium of the proposed stellar structure.
We conclude that the gravitational, hydrodynamic, pressure anisotropy and force due to quark matters enable the fulfillment of static equilibrium.

\section{Concluding Remarks}\label{con}
In the $5\mathcal{D}$ EGB gravity, we have primarily developed a hybrid star with baryonic and strange quark matter that is static, symmetric, and devoid of singularities and investigated its physical properties. We have selected a specific compact star, 4U 1538 - 52, for this theoretical inquiry, and throughout this paper, we will mathematically and graphically analyze our findings. For values of the coupling parameter $\alpha = 80, 90, 100, 110$ and $120;$ we plot all the physical characteristics.
In this paper, we assume the continuity of the metric functions $g_{tt}$, $g_{rr}$ and $\frac{\partial g_{tt}}{\partial r}$ across the boundary surface $r= \mathfrak{R}$ along with $p(r= \mathfrak{R})=0$ and match the interior metric with the exterior EGB version of the Schwarzschild metric in order to identify the constants included in the KB metric coefficients. 
We need to compare our findings to any available observational data. This kind of star might be conceivable in our nature because the model satisfies all necessary physical requirements and is horizon-free.
The important and startling conclusions described below have also been drawn by hypothetically constructing a new category of strange stars, which is supposed to have an interior domain and an outside region.\\
We outline our remarkable conclusions below:

\begin{itemize}

\item This work shows in detail how the MIT bag constant affects the various model parameters of candidates for anisotropic hybrid stars under $5\mathcal{D}$ EGB. The model was created based on the selection of the KB metric functions, which depend on the arbitrary constants $A, B$, and $C$. With the aid of the observed mass and radius of the star under consideration, we are able to successfully deduce the values of the aforementioned arbitrary constants and the values of $\gamma$ from the boundary conditions, which are listed in the reference Table \ref{table1}. Also, we noticed that $\gamma > 0$ always for arbitrary choice of $\alpha$.

\item It is crucial to create reliable physical models of compact stellar objects, and the graphical representations of $e^{\nu(r)}$ and $e^{\lambda(r)}$ in Fig.\ref{metric} clearly show that they are finite and non-singular over the radius of stars for varying $\alpha$.

\item The energy density $\rho(r)$ and pressure $p(r)$ inside the star remain continuous, positively finite, and exhibit a smooth decreasing tendency towards the surface for a range of $\alpha$ values, as is seen from Fig. \ref{rho}. We note that at the core $(r = 0)$, $\rho$ and $p$ are at their highest values, and at $r = R,$ radial pressure $p_r$ diminishes while positive values can be seen for transverse pressure and density. It is also an interesting fact that the presence of highly compact cores is indicated by the fact that density and pressure are both at their maximum levels at the core.
On the other hand, from Fig. \ref{rhoq} we can check the pressure and density profile due to quark matter and interestingly this figure indicates that the quark matter density $\rho_q$ and quark matter pressure $p_q$ are monotonically increasing towards the boundary surface of the star. In the background of other gravity systems, Bhar \cite{bhar2015new}, Abbas, and Nazar \cite{abbas2021hybrid} also discovered this unusual behavior of quark matter density. Also, we see that $\rho_q$ takes positive values and $p_q$ takes negative values throughout the stellar medium. On a galactic scale, the aforementioned findings might be interpreted in an intriguing way. Bag-model quark matter behaves as a dark matter if $p_q > 0$ and $\rho_q$ and $p_q$ correspond to the amount of matter in the galactic halo region. We now can assume that $\rho_q$ refers to the energy density produced by incorporating a region outside the halo large enough so that the ensuing decreased value causes $p_q < 0$ in equation (\ref{eos2}).

\item  In Table \ref{table2}, we evaluated the values of $\rho_c$, $\rho_s$ and $p_c$ which have the dependency on $B_g$. From this table, we have reached the conclusion that $\rho_c$, $\rho_s$, and $p_c$ remain positive throughout the stellar medium.

\item In Fig. \ref{grad}, the gradient components of matter-energy density and pressure exhibit a negative trend, going from zero to a negative region for different considerations of $\alpha$, whenever $r$ moves from the center to the boundary. The pressure and density gradients decrease monotonically throughout the stellar structure and drop at the center ($r=0$).

\item The evolution of compact strange star configurations is significantly influenced by physical factors like energy density and anisotropic stresses, i.e., $p_r$ and $p_t$, and the modeling of compact stars, however, also depends substantially on the pressure anisotropy, $\Delta$. Graphically, We have analyzed anisotropy as well as trace profile in Fig. \ref{delta} which confirms that the considered stellar model is viable.
 
\item It is clear from Fig. \ref{massu} that the mass function and compactness factor are consistent throughout the stellar zone and steadily rise with `r' and they reach their peak values at the star's border. At the right panel of this figure, we also show- how the compactness factor changes with the variation of coupling parameter $\alpha$. 

\item According to Fig. \ref{redshift}, the gravitational (or internal) redshift $(z_g)$ is minimum near the surface and greatest at the center, but surface redshift $(z_s)$ behaves exactly in the opposite fashion to $(z_g)$. In Table \ref{table3}, effective compactness and surface redshift function at $r = \mathfrak{R}$ are shown in tabular form in our study. 

\item As finding the equation of state, or a connection between pressure and density, is a critical step for the stability of our model, we have shown their profiles in Fig. \ref{eos}. Clearly, they are within the radiation era, that is, $0 < \omega_r, \omega_t < 1$. In addition, the graphs display how density influenced the radial and transverse pressures.
 
\item In order to keep the model consistent, we then looked at the stability mechanism. Firstly we have taken the causality criterion as well as Herrera's cracking concept and noticed that our suggested stellar model also meets the Causality criterion i.e., the square of the sound velocity $V^2$ lies within $(0,1)$ within the stellar body (From Fig. \ref{speed}), which makes the model physically stable and well-behaved.

\item In FIG. \ref{gama}, we examine the behavior of a crucial and important ratio between the two particulars given by the adiabatic index, $\Gamma$, moreover $\Gamma_r$, and $\Gamma_t$. In the graph, it can be seen that across the fluid sphere, both r and t take values bigger than $4/3$, ensuring that the stability criteria are fully satisfied. Furthermore, we can check $\Gamma_r, \Gamma_t > 4/3$ from Table \ref{table4}, where the numerical values of $\Gamma$ are shown.

\item For different values of $\alpha$, our resulting model satisfies all five energy criteria, namely the NEC, WEC, SEC, DEC, and TEC as shown graphically in Fig. \ref{Ener}. Additionally, these are consistently positive across the whole region of the star, supporting the feasibility of our suggested approach.

\item The forces associated with the TOV equation in Fig. \ref{force} show that the system is always in a state of equilibrium for any given value of $\alpha$. In this case, the gravitational force($F_g$) and the quark matter force($F_q$) are attracted while the hydrodynamic force found in ordinary matter($F_h$) and the anisotropy force($F_a$) are repulsive. 

As a result, a general and concluding observation is that our suggested model is a representation of a singularity-free, stable, and viable one that depicts a hybrid star formed of SQM and preferably fits strange star candidates to analyze their many physical aspects. We believe that our model might be able to play a larger role in the astrophysical scenario.

\end{itemize}

\section*{Author contributions}
\textbf{Akashdip Karmakar} performed mathematical analysis, computer code design for data analysis, numerical data analysis, and typed the draft. \textbf{Pramit Rej} contributed to conceptualization, computer code design for data analysis, original draft preparation, and overall supervision of the project. 

\section*{Acknowledgements} 
Pramit Rej is thankful to the Inter-University Centre for Astronomy and Astrophysics (IUCAA), Pune, Government of India, for providing Visiting Associateship.

\section*{Declarations}
\textbf{Funding:} The authors did not receive any funding in the form of financial aid or grant from any institution or organization for the present research work.\par
\textbf{Data Availability Statement:} The results are obtained
through purely theoretical calculations and can be verified analytically;
thus this manuscript has no associated data, or the data will not be deposited. \par
\textbf{Conflicts of Interest:} The authors have no financial interest or involvement that is relevant by any means to the content of this study.\par

\bibliography{hybrid_references}

\end{document}